\documentclass[a4paper,11pt]{article}
 \pdfoutput=1 % if your are submitting a pdflatex (i.e. if you have
\usepackage{jheppub} % for details on the use of the package, please
 \usepackage{epsfig}
\def\bma{\left( \begin{array} }
\def\ema{\end{array} \right)}
\newcommand{\vdel}{v_{\Delta}}
\newcommand{\delm}{\Delta M}

\preprint{KIAS-P12055}

\title{Vacuum Stability, Perturbativity, EWPD and Higgs-to-diphoton rate in Type II Seesaw Models}

\author{Eung Jin Chun,}
\author{Hyun Min Lee,}
\author{and Pankaj Sharma}

\affiliation{Korea Institute for Advanced Study\\
Heogiro 87, Dongdaemun-gu, Seoul 130-722, Korea}

\emailAdd{ejchun@kias.re.kr}
\emailAdd{hyun.min.lee@kias.re.kr}
\emailAdd{pankajs@kias.re.kr}

\abstract{
We study constraints from perturbativity and vacuum
stability as well as the EWPD in the type II seesaw model. As a
result, we can put stringent limits on the Higgs triplet couplings
depending on the cut-off scale. The EWPD tightly constrain the Higgs triplet
mass splitting to be smaller than 40 GeV.
Analyzing the Higgs-to-diphoton rate in the allowed parameter region,
we show a possible enhancement by up to 100 \% and
50 \% for the cut-off scale of 100 TeV and $10^{19}$ GeV,
respectively, if the doubly charged Higgs boson mass is as low as
100 GeV. }


\begin{document}
  \maketitle
%%%%%%%%%%%%%%%%%%%%%%%%%%%%%%%%%%%%%%%%%%%%%%%%%%%%%%%%%%%%%%%%%%%%%%%%%%%%%

\section{Introduction}

The discovery of the Higgs boson  at around 125 GeV \cite{higgs12} opened
a new era toward the Higgs precision test. It is essential for the LHC and future experiments
to determine how precisely the Higgs candidate follows the very prediction of
the Standard Model (SM), as new physics might enter here to
modify the SM Higgs property in various ways.

One of the motivations for new physics
beyond the SM comes from the smallness of neutrino masses whose origin can be attributed to
a new particle coupling to the lepton doublets of the SM.
In this paper, we consider the type II seesaw mechanism which introduces a Higgs triplet
whose vacuum expectation value (VEV) generates the neutrino masses and mixing \cite{type2}.
The Higgs sector of the type II seesaw contains four more bosons,
$H^{++}, H^+$ and $H^0/A^0$, in addition to the SM Higgs boson, $h$.
While the standard Higgs doublet generates the quark and charged lepton masses,
the Higgs triplet couples only to the lepton doublets generating the neutrino masses.
This mechanism leads to a peculiar prediction of a same-sign dilepton resonance,
$H^{++} \to l_\alpha^+ l_\beta^+$, which is being searched at the LHC \cite{cmsH++}.
As the Higgs triplet Yukawa matrix is proportional to the neutrino mass matrix,
the observation of the flavor structure of the same-sign dilepton final states allows us
to determine the neutrino mass pattern at colliders \cite{chun03}.

\medskip

Other interesting features of the type II seesaw come from the Higgs boson sector.
The Higgs triplet couplings can change drastically the stability of the
SM electroweak vacuum  \cite{vacuumstability} so they are quite constrained.
Furthermore, in the limit of tiny lepton Yukawa couplings of the Higgs triplet,
the triplet self couplings are also constrained by perturbativity as they tend to blow up rapidly.
Considering the perturbativity and absolute vacuum stability conditions up to the Planck scale,
we will see that perturbativity keeps a triplet self coupling, denoted by $\lambda_2$,
 smaller than $0.25$ and then
vacuum stability requires all the other couplings to be smaller than $0.5$.
If a lower instability scale is taken, such a stringent limit can of course be relaxed, but not too much.

Another important constraint can be deduced from the electroweak precision data (EWPD) \cite{melfo11}.
Note that one of the couplings between the Higgs triplet and doublet,
 denoted by $\lambda_5$, induces mass splitting $\Delta M$
among the triplet components \cite{chun03}. The EWPD
turn out to put a strong limit of $|\Delta M| \lesssim 40$ GeV allowing only a narrow range of
$\lambda_5$ depending on the Higgs triplet mass
 when the triplet VEV is taken to be tiny enough so that its
 tree-level contribution to $\Delta \rho$ is neglected.

As noted in \cite{arhrib11,kanemura12,akeroyd12},  the SM Higgs boson decay
$h\to \gamma\gamma$ can be significantly modified through one-loop diagrams
involving the charged Higgs bosons, in particular, $H^{++}$, if quartic
couplings mixing with the SM Higgs are large and the triplet mass is small.
Recently it is of a particular interest to look for plausible models accommodating
the enhanced Higgs-to-diphoton rate that appeared in the current LHC data \cite{xx-hgg}.
Whether or not such a deviation is confirmed by a further LHC search, the precise measurement of
the diphoton rate will place an important restriction on the type II seesaw model.
In our analysis, we show how much the $h\to \gamma\gamma$ rate can deviate from the SM prediction
after restricting ourselves to the model parameter space allowed by
the perturbativity and vacuum stability conditions as well as the EWPD constraint, which has
not been considered properly in the previous studies \cite{arhrib11,kanemura12,akeroyd12}.
 As expected, the result strongly depends on the assumed instability scale as well as the Higgs triplet mass.
For our analysis, we will take the instability scale at 100 TeV, $10^{10}$ GeV and $10^{19}$ GeV.
The doubly charged Higgs boson mass is taken to be as low as 100 GeV although it can be even smaller
depending on the assumed decay channels of the triplet components.

All the collider searches for the doubly charged Higgs boson so far look for the clean signal of
$H^{++} \to l^+_\alpha l^{+}_\beta $.  The previous results from LEP \cite{lep2}
and Tevatron \cite{cdf} put lower limits on the charged Higgs boson mass at around 100 GeV
assuming the 100 \% branching fraction for the $H^{++}$ decay to some specific lepton flavours.
The current CMS analysis includes more realistic dilepton decay channels reflecting
the allowed neutrino mass pattern and puts the  lower limit of 333 GeV -- 408 GeV depending
on the chosen benchmark points and decay scenarios \cite{cmsH++}.
But this limit is applicable when the same-sign dilepton
branching ratio is again 100\%. In fact, the doubly charged Higgs can have three types of
decay channels: $H^{++} \to l^+_\alpha l^+_\beta$, $W^+ W^+$ and $H^+ W^+$,
one of which can dominate
depending on the model parameters \cite{chun1206}.
If one considers the triplet VEV larger than about $10 ^{-4}$ GeV,
$H^{++}$ can decay dominantly to $W^+ W^+$  reducing the dilepton branching ratio and thus the lower limit on the doubly charged Higgs boson mass. It may be worthwhile to make more
serious studies to search for the $H^{++} \to W^+ W^+$ signal in the upcoming 14 TeV LHC \cite{perez,Chiang:2012dk}. The worst scenario for the doubly charged Higgs search would be when there
is a sizable mass gap among the triplet components and the doubly charged state is the heaviest
(for $\lambda_5<0$ as we will see).
In this case, the produced doubly charged Higgs boson follows the decay chain: $H^{++} \to H^+ W^{(*)} \to H^0/A^0 W^{(*)} W^{(*)} \to \nu \nu W^{(*)} W^{(*)}$
and thus the triplet can be completely missed.

\medskip

This paper is organized as follows.
After introducing the type II seesaw model
with the model parameters and notations for our analysis in Section 2, we will
find constraints on the Higgs couplings by applying the conditions for the perturbativity
and vacuum stability, and the renormalization group equations at one-loop in Section 3.
Then, additional
restrictions from the EWPD will be obtained in Section 4. We will calculate the modified
Higgs-to-diphoton rate due to the $H^{++}$ and $H^+$ contribution
depending on the allowed ranges of the Higgs triplet couplings in
Section 5. The results of our combined analysis will be presented and conclusions
will be drawn in Section 6.

\section{Higgs couplings in type II seesaw}

When the Higgs sector of the Standard Model is extended with a
$Y=2$  $SU(2)_L$ scalar triplet $\Delta$ in addition to a
SM-Higgs doublet $\Phi$, the gauge-invariant Lagrangian is written
as
\begin{eqnarray}
\mathcal L=\nonumber\left(D_\mu\Phi\right)^\dagger
\left(D^\mu\Phi\right)  + \mbox{Tr}
\left(D_\mu\Delta\right)^\dagger\left(D^\mu\Delta\right) -\mathcal
L_Y - V(\Phi,\Delta)
\end{eqnarray}
where the leptonic part of the Lagrangian required to generate
neutrino masses is
\begin{equation} \label{leptonYuk}
\mathcal L_Y= f_{\alpha\beta}L_\alpha^T Ci\tau_2\Delta L_{\beta} +
\mbox{H.c.}
\end{equation}
and the scalar potential is
\begin{eqnarray}\label{Pot}
V(\Phi,\Delta)&=&\nonumber m^2\Phi^\dagger\Phi +
\lambda_1(\Phi^\dagger\Phi)^2+M^2\mbox{Tr}(\Delta^\dagger\Delta)\\\nonumber
&+&\lambda_2\left[\mbox{Tr}(\Delta^\dagger\Delta)\right]^2
+2\lambda_3\mbox{Det}(\Delta^\dagger\Delta)
+\lambda_4(\Phi^\dagger\Phi)\mbox{Tr}(\Delta^\dagger\Delta)\\
&+&\lambda_5(\Phi^\dagger\tau_i\Phi)\mbox{Tr}(\Delta^\dagger\tau_i\Delta)
+\left[\frac{1}{\sqrt{2}}\mu(\Phi^Ti\tau_2\Delta^\dagger\Phi)+\mbox{H.c.}\right].
\end{eqnarray}
Here used is the $2\times 2$ matrix representation of $\Delta$:
\begin{equation}
\Delta= \left( \begin{array}{cc}
\Delta^+/\sqrt{2}  & \Delta^{++} \\
\Delta^0       & -\Delta^+/\sqrt{2} \end{array} \right) .
\end{equation}
Upon the electroweak symmetry breaking with $\langle
\Phi^0\rangle=v_0/\sqrt{2}$, the $\mu$ term in Eq.~(\ref{Pot}) gives rise to the
vacuum expectation value of the triplet $\langle
\Delta^0\rangle=v_\Delta/\sqrt{2}$ where $\vdel\approx \mu
v_0^2/\sqrt{2}M^2$. We will assume $\mu$ is real positive without loss of generality.
From the leptonic Yukawa coupling (\ref{leptonYuk}), one can get the neutrino mass matrix
\begin{equation}
 M^\nu_{\alpha\beta} = f_{\alpha\beta}\, \xi \,v_0 ,
\end{equation}
where $\xi\equiv v_\Delta/v_0$. The observed neutrino mass of order $0.1$ eV requires
$|f_{\alpha\beta}\,\xi| \sim 10^{-12}$. Considering this relation, we will
assume $|f_{\alpha\beta}| \ll 1$ and $|\xi|\ll 1$ throughout this work.
Let us remind that the measurement of $\rho \equiv M_W^2/(M_Z^2 c_W^2 )\approx 1$ puts the bound
$\xi \lesssim 10^{-2}$. Since we further take the region of $|\xi| \ll 10^{-2}$ in our analysis,
some of the effects with a largest possible value of $\xi$ \cite{shafi08,kanemura12} can be safely neglected.

After the electroweak symmetry breaking, there are five physical
massive bosons denoted by $H^{\pm\pm}$, $H^\pm$,
$H^0$, $A^0$, $h^0$. Under the condition of $|\xi|\ll 1$,
the first five states are mainly from
the triplet scalar and the last from the doublet scalar. For the
neutral pseudoscalar and charged scalar parts,
\begin{eqnarray}
 \phi^0_I = G^0 - 2 \xi A^0 \;, \qquad
 \phi^+ = G^+ + \sqrt{2} \xi H^+  \nonumber\\
 \Delta^0_I= A^0 + 2 \xi G^0 \;, \qquad
 \Delta^+= H^+ - \sqrt{2} \xi G^+
\end{eqnarray}
where $G^0$ and $G^+$ are the Goldstone modes, and
for the neutral scalar part,
\begin{eqnarray}
 \phi^0_R &=& h^0 - a \xi \, H^0 \,, \nonumber\\
 \Delta^0_R&=& H^0 + a \xi \, h^0
\end{eqnarray}
where $ a = 2 + (4\lambda_1-\lambda_4-\lambda_5) v_0^2/(M^2_{H^0}-M^2_{h^0}) $.
The masses of the Higgs bosons essentially from the triplet are
\begin{eqnarray} \label{massD}
 M^2_{H^{\pm\pm}} &=& M^2 + {\lambda_4 -\lambda_5 \over 2} v_0^2
 \nonumber\\
 M^2_{H^{\pm}} &=& M_{H^{\pm\pm}}^2 + {\lambda_5 \over 2} v^2_{0}
 \nonumber\\
 M^2_{H^0, A^0} &=&  M^2_{H^{\pm\pm}} +
    {\lambda_5 } v^2_{0} \,,
\end{eqnarray}
neglecting small contributions from $v_\Delta$.
The mass of $h^0$ is
given by $m_{h^0}^2=2\lambda_1 v_0^2$ as usual.

Eq.~(\ref{massD}) tells us that the mass splitting,
$\delm \equiv  M_{H^{\pm}} - M_{H^{\pm\pm}}$, is driven by the coupling $\lambda_5$
which affects also the EWPD and the Higgs-to-diphoton rate.
Recall that depending upon the sign of the coupling $\lambda_5$,
there are two mass hierarchies among the triplet components:
$M_{H^{\pm\pm}}>M_{H^\pm}>M_{H^0,A^0}$ for $\lambda_5<0$; or
$M_{H^{\pm\pm}}<M_{H^\pm}<M_{H^0,A^0}$ for $\lambda_5> 0$ \cite{chun03}.
The charged Higgs boson as light as 100 GeV ($M_{H^{\pm\pm}}$ or $M_{H^\pm} = 100$ GeV)
can evade the CMS search if the decay channels of $H^{\pm\pm} \to H^\pm W^*$ and
$H^\pm \to H^0/A^0 W^*$ are the dominant modes allowed by a sizable $\lambda_5$
in the first case, or if $H^{\pm\pm}$ decays dominantly to $W^\pm W^\pm$ with
$|\xi| \gg |f_{ij}|$ in the second case.

\section{Vacuum stability and perturbativity}

The scalar potential (\ref{Pot}) contains seven free parameters:
$\lambda_i\, (i=1\ldots 5$), $v_\Delta$ and $M_{H^{++}}$.
Rather stringent constraints on these
parameters can be readily obtained by the theoretical requirements of perturbativity
and vacuum stability.  A detailed study of the scalar potential has been performed
in \cite{arhrib1105}. The vacuum stability conditions on the scalar couplings
$\lambda_i$   are as follows:
\begin{eqnarray} \label{vcst}
&&\lambda_1>0, \quad \lambda_2>0,\quad \lambda_2 + {1\over2} \lambda_3 >0
\\
&&\lambda_4 \pm \lambda_5 + 2 \sqrt{\lambda_1\lambda_2} >0, \quad
\lambda_4 \pm \lambda_5 + 2 \sqrt{\lambda_1(\lambda_2+{1\over2}\lambda_3)} >0.
\nonumber
\end{eqnarray}
Apart from these conditions, we will put the perturbativity conditions:
$|\lambda_i| \leq \sqrt{4\pi}$.

We will take the absolute stability condition\footnote{Imposing metastability \cite{metastability}
instead of absolute stability would lead to a wider parameter space but we don't consider
this possibility in our work.} that
all these constraints must remain
true up to the scale where the theory is supposed to be valid.
Henceforth, we study the
renormalization group (RG) evolution of these scalar couplings ($\lambda_i$'s),
EW-gauge couplings $g_2$, $g^\prime$, strong coupling $g_3$ and top-Yukawa coupling $y_t$
up to the cut-off scale at the one-loop
level. The RG evolution of the type II seesaw model has been studied in \cite{schmidt07}.
The one-loop RG equations relevant for our analysis are as below:
\begin{eqnarray} \label{rge}
16\pi^2 {d g^\prime \over d t} &=& {47\over6} g^{\prime 3},\quad
16\pi^2 {d g_2 \over d t} = -{5\over2} g_2^{3},\quad
16\pi^2 {d g_3 \over d t} = -7 g_3^{3};
\\
16\pi^2 {d y_t \over d t} &=&
y_t ( {9\over2} y_t^2 -{17\over 12} g^{\prime 2} -{9\over4} g_2^2-8 g_3^2)
\nonumber\\
16\pi^2 {d\lambda_1 \over d t} &=&
24 \lambda_1^2 + \lambda_1 (-9g_2^2-3g^{\prime 2}+12 y_t^2)
+{3\over4} g_2^4 +{3\over8} (g^{\prime 2}+g_2^2)^2 - 6 y_t^4 + 3 \lambda_4^2 + 2\lambda_5^2
\nonumber\\
16\pi^2 {d\lambda_2 \over d t} &=&
\lambda_2(-12 g^{\prime 2}-24g_2^2) +6 g^{\prime 4} + 9 g_2^4 + 12 g^{\prime 2} g_2^2
+28 \lambda_2^2 + 8 \lambda_2\lambda_3+4\lambda_3^2+2\lambda_4^2+2\lambda_5^2
\nonumber\\
16\pi^2 {d\lambda_3 \over d t} &=&
\lambda_3(-12 g^{\prime 2}-24g_2^2) + 6 g_2^4 -24 g^{\prime 2} g_2^2
+6 \lambda_3^2 + 24\lambda_2\lambda_3 - 4\lambda_5^2
\nonumber\\
16\pi^2 {d\lambda_4 \over d t} &=&
\lambda_4(-{15\over2} g^{\prime 2} - {33\over2}g_2^2) + {9\over5} g^{\prime 4} + 6 g_2^4
+\lambda_4 (12\lambda_1 + 16\lambda_2 + 4\lambda_3 + 4\lambda_4 + 6 y_t^2) + 8 \lambda_5^2
\nonumber\\
16\pi^2 {d\lambda_5 \over d t} &=&
\lambda_4(-{15\over2} g^{\prime 2} - {33\over2}g_2^2) + 6 g^{\prime 2} g_2^2
+\lambda_5 (4\lambda_1 + 4\lambda_2 - 4\lambda_3 + 8\lambda_4 + 6 y_t^2),
\nonumber
\end{eqnarray}
where $t\equiv \ln(\mu/M_t)$ and the contributions
from the neutrino Yukawa couplings, $f_{\alpha\beta}$, are neglected.

\begin{figure}[ht]
\begin{center}\hspace{-1.1cm}
\includegraphics[scale=0.24]{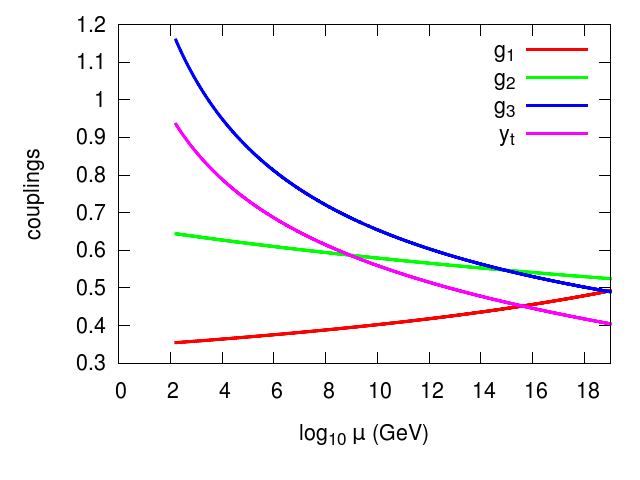}\hspace{-0.2cm}
\includegraphics[scale=0.24]{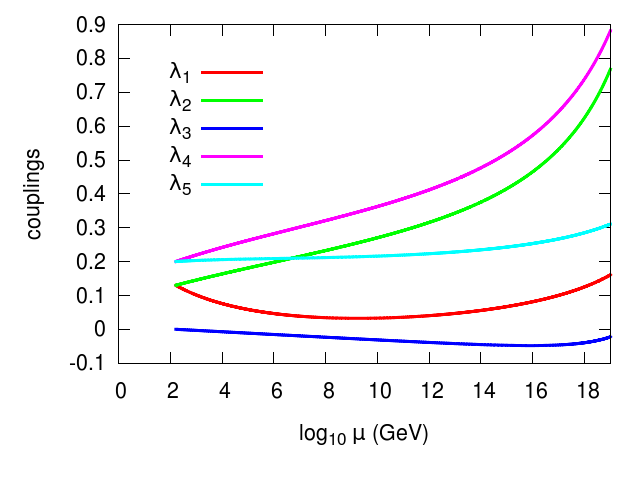}\hspace{-0.2cm}
\includegraphics[scale=0.24]{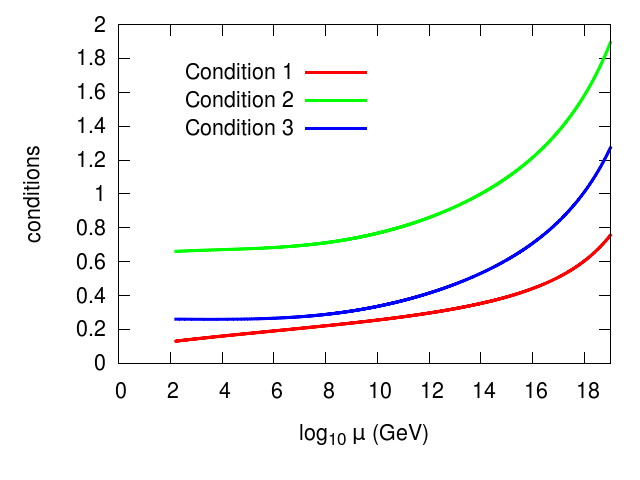}
\end{center}
\caption{RG evolution of couplings and vacuum stability conditions.} \label{fig:RGE}
\end{figure}

In Fig.~\ref{fig:RGE}, we show an example of the RG running of the couplings
which maintain the perturbativity and vacuum stability up to the Planck scale.
In the rightmost panel, the three vacuum stability conditions;
(1) $\lambda_2 + {1\over2} \lambda_3 >0$,
(2) $\lambda_4 - \lambda_5 + 2 \sqrt{\lambda_1\lambda_2} >0$, and
(3) $\lambda_4 - \lambda_5 + 2 \sqrt{\lambda_1
(\lambda_2+{1\over2}\lambda_3)} >0$ are
presented.
 Note that the Higgs doublet self-coupling $\lambda_1$ decreases initially due to the top Yukawa
coupling as in the SM, but it turns around to increase at a certain point with the aid of
other increasing couplings.
For our numerical analysis, we use $M_t = 173$ GeV, $m_t(M_t)=164$ GeV,
$m_h=125$ GeV and thus $\lambda_1(M_t)= m_h^2/2 v_0^2=0.129$ and $y_t(M_t)=\sqrt{2} m_t/v_0=0.938$.

\section{Constraints from EWPD}

In this section, we study the contributions of the Higgs triplet
 to the EWPD observables, also known as the oblique parameters. In \cite{lavoura93},
the contribution of a scalar multiplet of arbitrary weak isospin and weak hypercharge
 to the $S$, $T$ and $U$ parameters has been calculated.
We present here the expressions for the specific case of the Higgs triplet model:
\begin{eqnarray} \label{stu}
S &=& -{1\over3\pi} \ln{m_{+1}^2 \over m_{-1}^2}
-{2\over\pi} \sum_{T_3=-1}^{+1} (T_3 - Q s_W^2)^2 \,
\xi\left({m_{T_3}^2\over m_Z^2}, {m_{T_3}^2\over m_Z^2}\right) \\
T &=& {1\over 16\pi c_W^2 s_W^2} \sum_{T_3=-1}^{+1} \left(2-T_3(T_3-1)\right)\,
\eta\left({m_{T_3}^2\over m_Z^2}, {m_{T_3-1}^2\over m_Z^2}\right) \nonumber\\
U &=& {1\over6\pi} \ln{m_{0}^4 \over m_{+1}^2 m_{-1}^2}
+{1\over\pi} \sum_{T_3=-1}^{+1} \left[ 2(T_3 - Q s_W^2)^2\,
\xi\left({m_{T_3}^2\over m_Z^2}, {m_{T_3}^2\over m_Z^2}\right) \right.\nonumber\\
&& \left. ~~~~~~~~~~~~~~~~~~~~~~~~~~~~
-(2-T_3(T_3-1))\, \xi\left({m_{T_3}^2\over m_W^2}, {m_{T_3}^2\over m_W^2}\right)\right] \nonumber
\end{eqnarray}
where $m_{+1,0,-1} = M_{H^{++},H^+,H^0}$ and the functions $\xi(x,y)$ and $\eta(x,y)$ are defined by
\begin{eqnarray}
 \xi(x,y) &=& {4\over9} -{5\over12}(x+y) +{1\over6}(x-y)^2\\
 &&+{1\over4}\left[ x^2-y^2-{1\over3}(x-y)^3
 -{x^2+y^2\over x-y}\right] \ln{x\over y} -{1\over12} d(x,y) f(x,y)\nonumber \\
 d(x,y) &=& -1+2(x+y)-(x-y)^2 \nonumber\\
 f(x,y) &=& \begin{cases}
 -2 \sqrt{d(x,y)}\left[ \arctan{x-y+1\over\sqrt{d(x,y)}} -  \arctan{x-y-1\over\sqrt{d(x,y)}}\right]
 \quad\mbox{for}\quad d(x,y)>0 \cr
  \sqrt{-d(x,y)} \ln\left[ {x+y-1+\sqrt{-d(x,y)} \over x+y-1-\sqrt{-d(x,y)}}\right]
 \quad\mbox{for}\quad d(x,y)\leq0 \cr \end{cases} \nonumber\\
 \eta(x,y) &=& x+y - {2 xy\over x-y} \ln{x\over y} \nonumber
\end{eqnarray}
Adopting the most recent fit results for the allowed regions of the $S$, $T$ and $U$ presented
in \cite{Baak:2012kk}, we use the following values for the SM fit of the oblique parameters:
\begin{align}
 S_{\rm best\;fit}&= 0.03 \,  ,&&\sigma_S=0.10 \, , \\
 T_{\rm best\;fit}&= 0.05 \,  ,&&\sigma_T=0.12 \, , \nonumber\\
 U_{\rm best\;fit}&= 0.03 \,  ,&&\sigma_U=0.10 \, , \nonumber
\end{align}
As the $S$, $T$ and $U$ are not independent quantities, there is a correlation among these quantities.
The correlation coefficients are given by
\begin{equation}
\rho_{ST} = 0.89,\quad \rho_{SU} = -0.54,\quad \rho_{TU}=-0.83
\end{equation}
The contour allowed by the EWPD at a given confidence level $CL$ is then determined by
\begin{align}
 \begin{pmatrix}
  S-S_{\rm best\;fit}\\
  T-T_{\rm best\;fit}\\
  U-U_{\rm best\;fit}
 \end{pmatrix}^T
  \begin{pmatrix}
    \sigma_S \sigma_S       &    \sigma_S \sigma_T \rho_{ST} &  \sigma_S \sigma_U \rho_{SU}\\
    \sigma_S \sigma_T \rho_{ST} &    \sigma_T \sigma_T       &  \sigma_T \sigma_U \rho_{TU}\\
    \sigma_U \sigma_S \rho_{US} &    \sigma_U \sigma_T \rho_{TU} &  \sigma_U \sigma_U
 \end{pmatrix}^{-1}
  \begin{pmatrix}
  S-S_{\rm best\;fit}\\
  T-T_{\rm best\;fit}\\
  U-U_{\rm best\;fit}
 \end{pmatrix} = -2 \ln(1-CL)\;.
\end{align}
\label{STUFit}

\begin{figure}[ht]
\begin{center}\hspace{-1.1cm}
\includegraphics[scale=0.4]{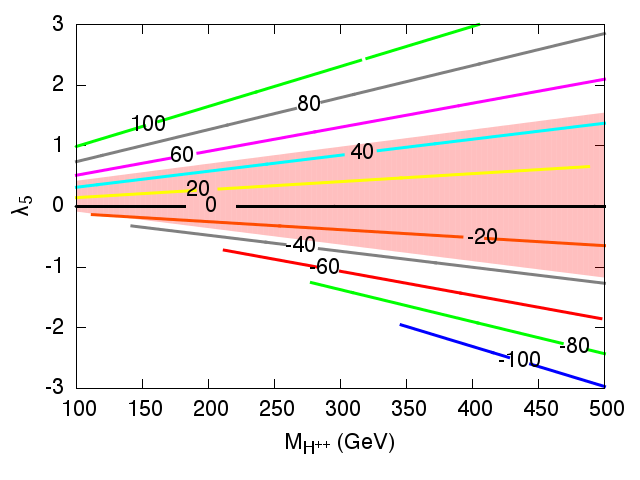}\hspace{-0.2cm}
\end{center}
\caption{Allowed parameter space in the $M_{H^{++}}$--$\lambda_5$ plane.
The contours represent
the allowed values of mass splitting, $\Delta M\equiv M_{H^+}-M_{H^{++}}$,
in the unit of GeV. The shaded
band denotes the 99\%  CL region satisfying the EWPD constraint.}\label{ewpd:mhpp-l5}
\end{figure}

In Fig.~\ref{ewpd:mhpp-l5}, we show the allowed parameter space in the $M_{H^{++}}$--$\lambda_5$
plane consistent with the EWPD. The shaded region shows the EWPD constraint at 99\% CL.
The contour lines show the mass splitting, $\Delta M\equiv M_{H^+}-M_{H^{++}}$, from which
one can see that the mass splitting is tightly constrained to be within $|\Delta M| \lesssim 40$
GeV independently of the doubly charged Higgs mass.

Note that this conclusion can be changed if a relatively large triple VEV,
$v_\Delta\gtrsim 1 \,{\rm GeV}$ ($\xi\sim 0.01$), is assumed \cite{kanemura12} in which case a sizable tree-level $\delta\rho$
contribution coming from the triplet VEV can be cancelled out 
by loop contributions with a large mass
splitting among the triplet components to satisfy the EWPD constraints.

\section{Higgs triplet contribution to  $h\to \gamma\gamma$}

Having studied the consistency conditions on the model parameters,
we now analyze their impact on the Higgs boson decay to two photons.
In the type II seesaw model, the Higgs-to-diphoton decay rate gets a sizable contribution from the charged Higgs bosons,
$H^{++}$ and $H^+$, which can lead to a constructive or destructive interference
with the SM contribution from the top quark and weak gauge boson.
Summing up all the contributions, one gets the following Higgs-to-diphoton rate
\cite{djouadi05}:
\begin{eqnarray}
\Gamma(h\to \gamma\gamma) &=& {G_F \alpha^2 m_h^3\over 128\sqrt{2}\pi^3}
\left| \sum_f N_c Q_f^2\, g^h_{ff} A^h_{1/2}(x_f) + g^h_{WW} A^h_1(x_W) \right.\\
&& \left. +  g^h_{H^+ H^-} A^h_0(x_{H^+})
+ 4g^h_{H^{++} H^{--}} A^h_0(x_{H^{++}}) \right|^2 \nonumber
\end{eqnarray}
where $x_i = m_h^2/4 m_i^2$ and the functions are
\begin{eqnarray}
A^h_{1/2}(x) &=& 2x^{-2}[x + (x-1) f(x)] \\
A^h_{1}(x) &=& -x^{-2}[2x^2 + 3 x + 3 (2x-1) f(x)] \nonumber\\
A^h_0(x) &=& -x^{-2}[x-f(x)] \nonumber\\
\quad\mbox{where}&&\!\!\!\! f(x) =
 \begin{cases}
 \arcsin^2\sqrt{x} \quad\mbox{for}\quad x\leq 1 \cr
 - {1\over 4} \left[ \ln{ 1+\sqrt{1-x^{-1}} \over 1-\sqrt{1-x^{-1}}} - i\pi\right]^2
\quad\mbox{for}\quad x>1 \cr \end{cases} \nonumber
\end{eqnarray}
The Higgs couplings are
$g^h_{ff} =1$ for the top and $g^h_{WW}=1$, whereas the Higgs triplet couplings are
\begin{equation}
 g^h_{H^+ H^+} = {\lambda_4 \over 2} {  v_0^2 \over M_{H^+}^2} , \quad\mbox{and}\quad
 g^h_{H^{++} H^{++}} = {\lambda_4 -\lambda_5 \over 2} { v_0^2 \over M_{H^{++}}^2} \,.
\end{equation}
Since the SM contribution amounts to about $-6.5$ in the amplitude, negative values
of $\lambda_4$ and $\lambda_4 -\lambda_5$ can make a constructive interference to enhance the diphoton rate.
As we will see in the next section, however,
the vacuum stability condition strongly disfavors negative $\lambda_4$
and $\lambda_4 - \lambda_5$ and allows more parameter region leading to
a destructive interference to reduce the diphoton rate.

\begin{figure}[ht]
\begin{center}\hspace{-0.9cm}
\includegraphics[scale=0.42]{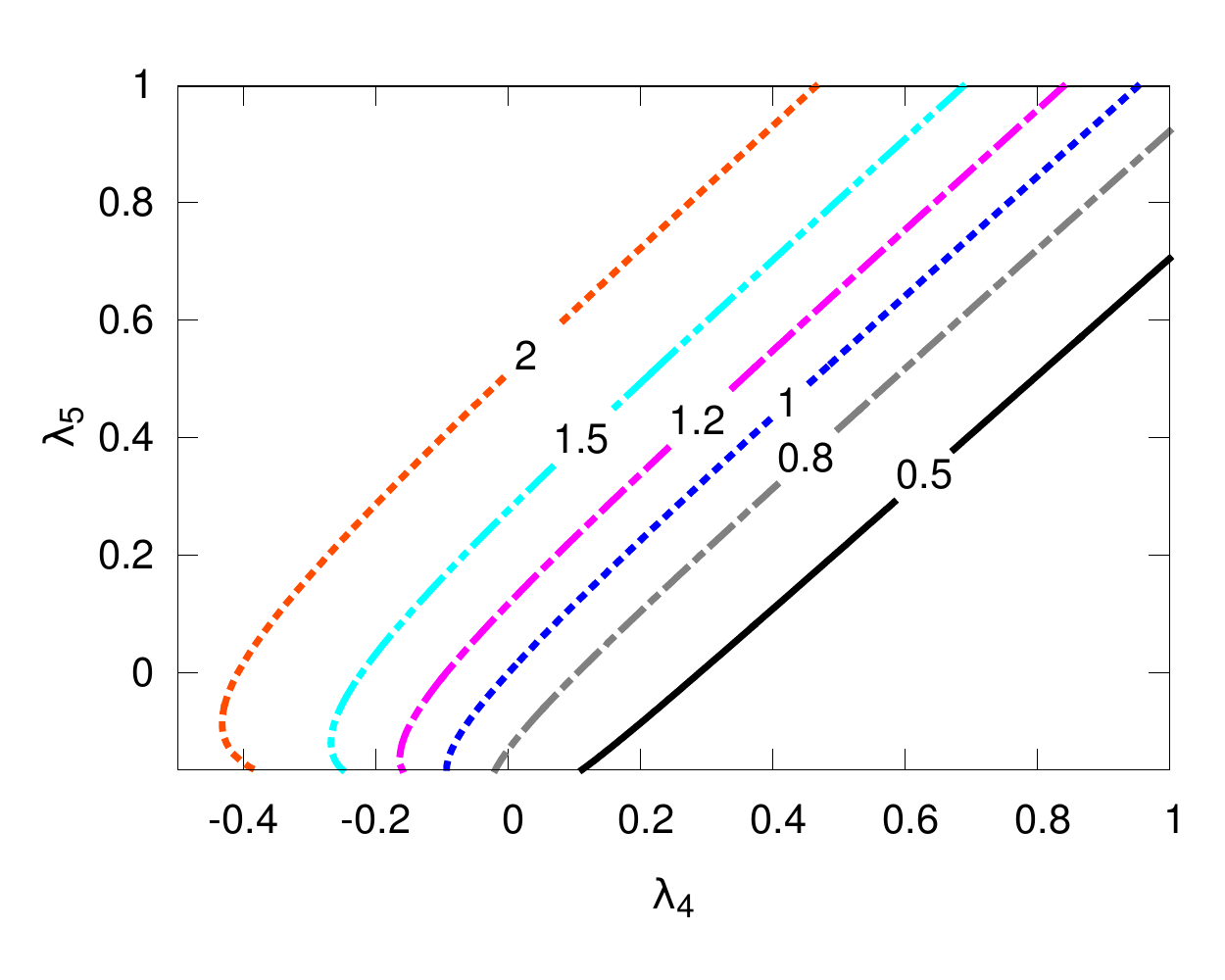}\hspace{-0.2cm}
\includegraphics[scale=0.42]{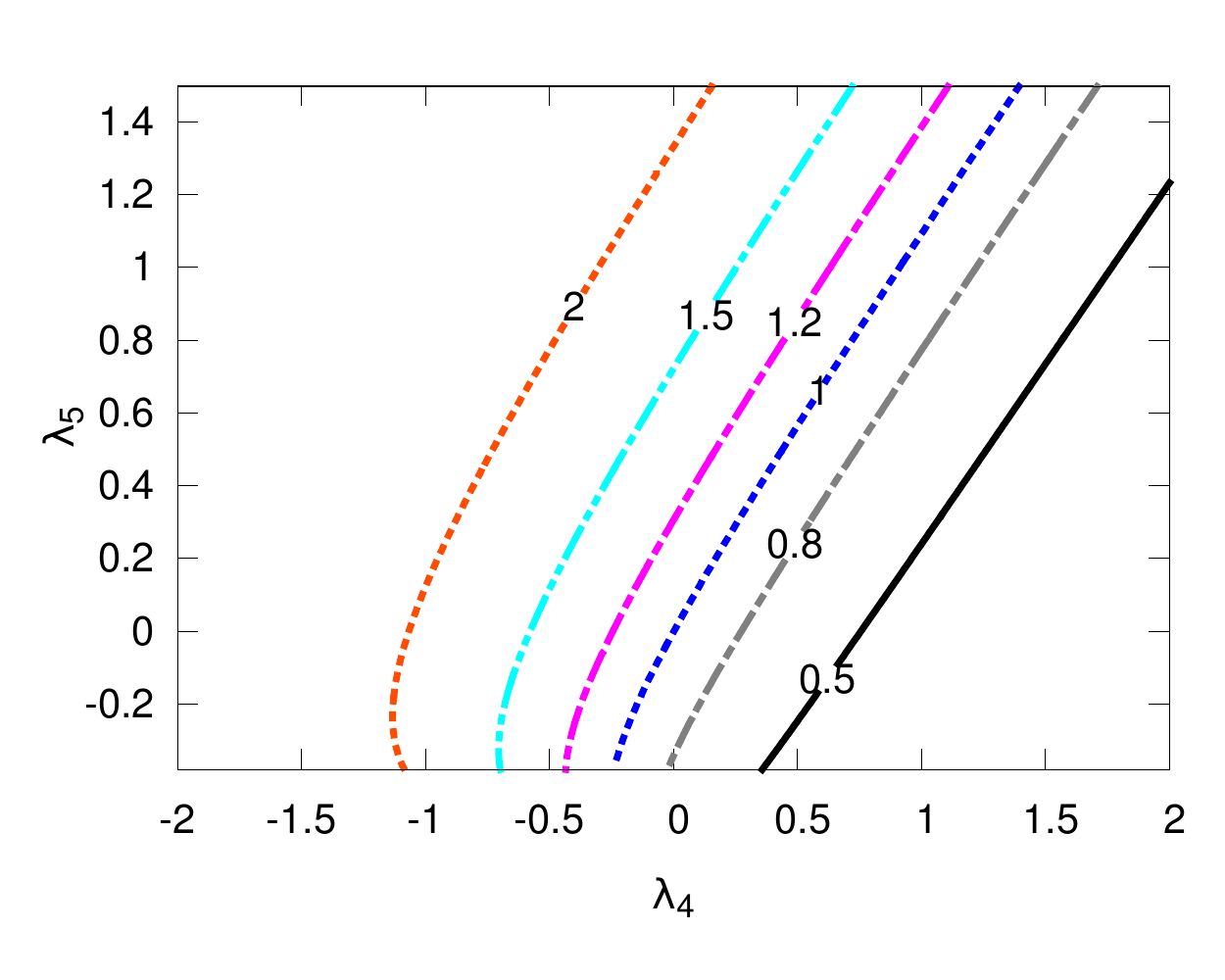}\hspace{-0.2cm}
\includegraphics[scale=0.42]{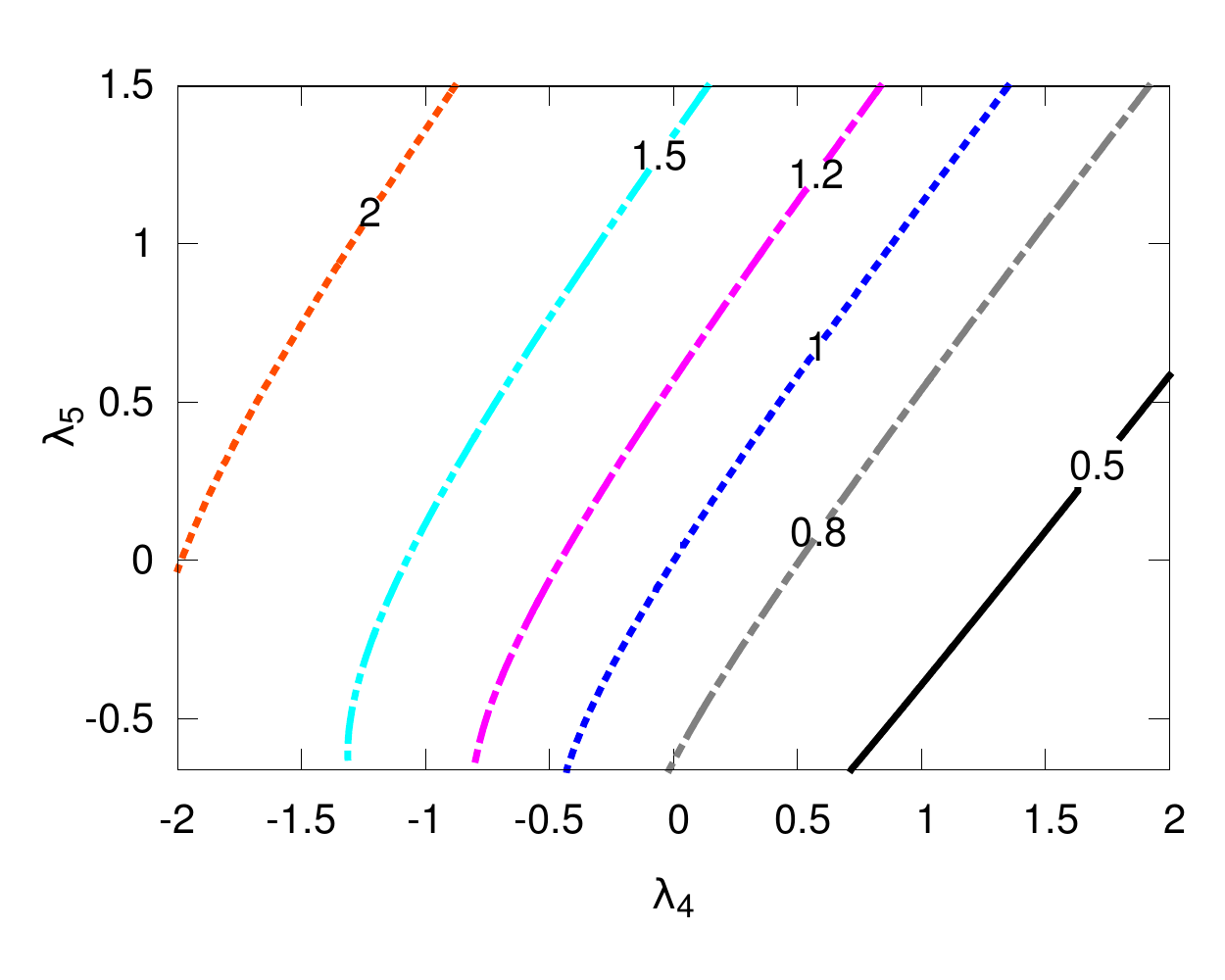}
\end{center}
\caption{The $R_{\gamma\gamma}$ contours in the $\lambda_4$--$\lambda_5$ plane
for $M_{H^{++}}=100$ GeV (left), 150 GeV (middle)
and 200 GeV (right).}\label{Rgg}
\end{figure}

Fig.~\ref{Rgg} shows the contour lines of $R_{\gamma\gamma} \equiv \Gamma(h\to \gamma\gamma)/
\Gamma(h\to \gamma\gamma)|_{\rm SM}$ in the $\lambda_4$--$\lambda_5$ plane for the doubly charged
Higgs masses, $M_{H^{++}}=$ 100, 150, and 200 GeV.
In the region with $\lambda_5<0$, the singly charged Higgs is lighter than the
doubly charged Higgs and its constructive contribution becomes more sizable so that the contour lines
start to bend for a certain value of $\lambda_5$. The contour lines are cut at the $\lambda_5$ values beyond which
the mass-squared values of the neutral components of triplet become negative.
Note that for $M_{H^{++}}=$ 100 GeV, 150 GeV and 200 GeV, the positivity of mass-squared values of the
neutral Higgs requires $\lambda_5\geq -0.165,~-0.38$ and $-0.66$ respectively.
In the next section, the constraints derived in the previous two sections are combined and overlayed
with the $R_{\gamma\gamma}$ contours. We will see that the EWPD constraint derived in Fig.~\ref{ewpd:mhpp-l5}
restricts $\lambda_5$ to a smaller region than in Fig.~\ref{Rgg}.

\section{Results and summary}

\begin{figure}[t]
\begin{center}

\hspace{-1.4cm}
\includegraphics[scale=0.48]{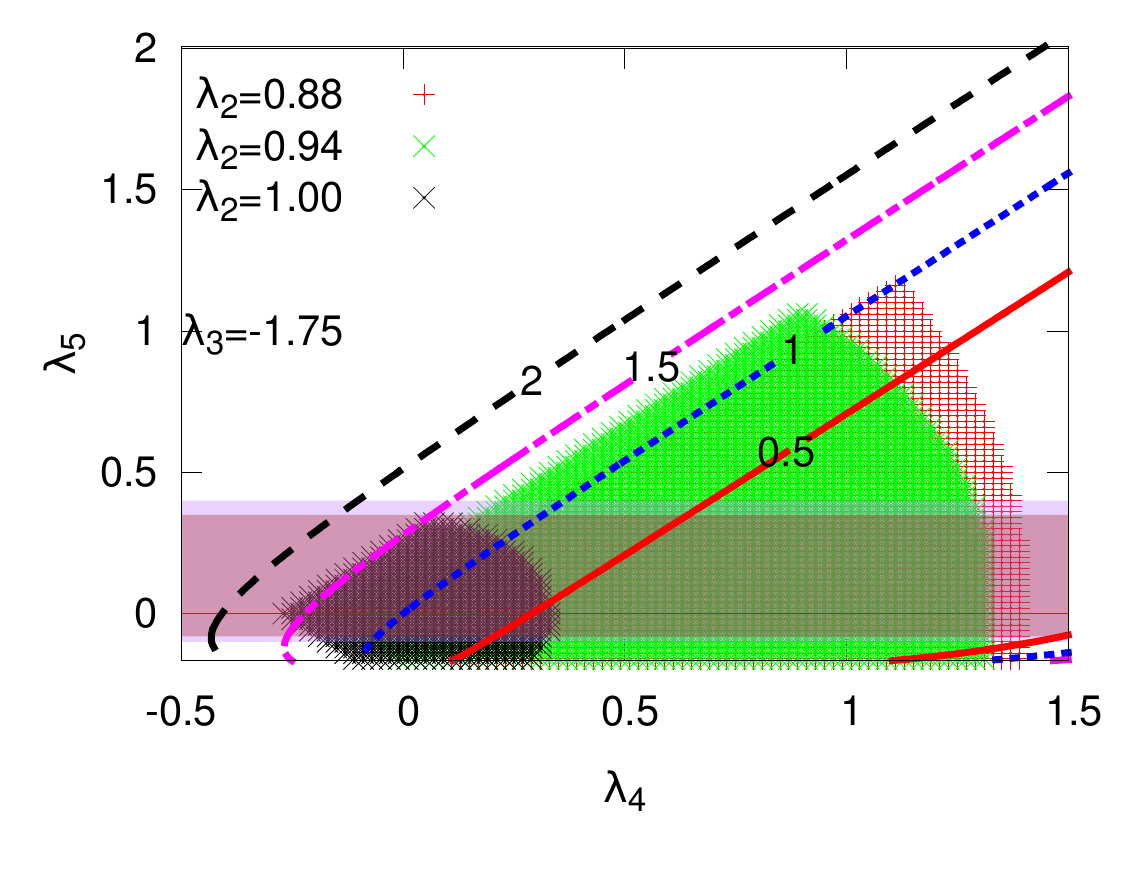}\hspace{-0.2cm}
\includegraphics[scale=0.48]{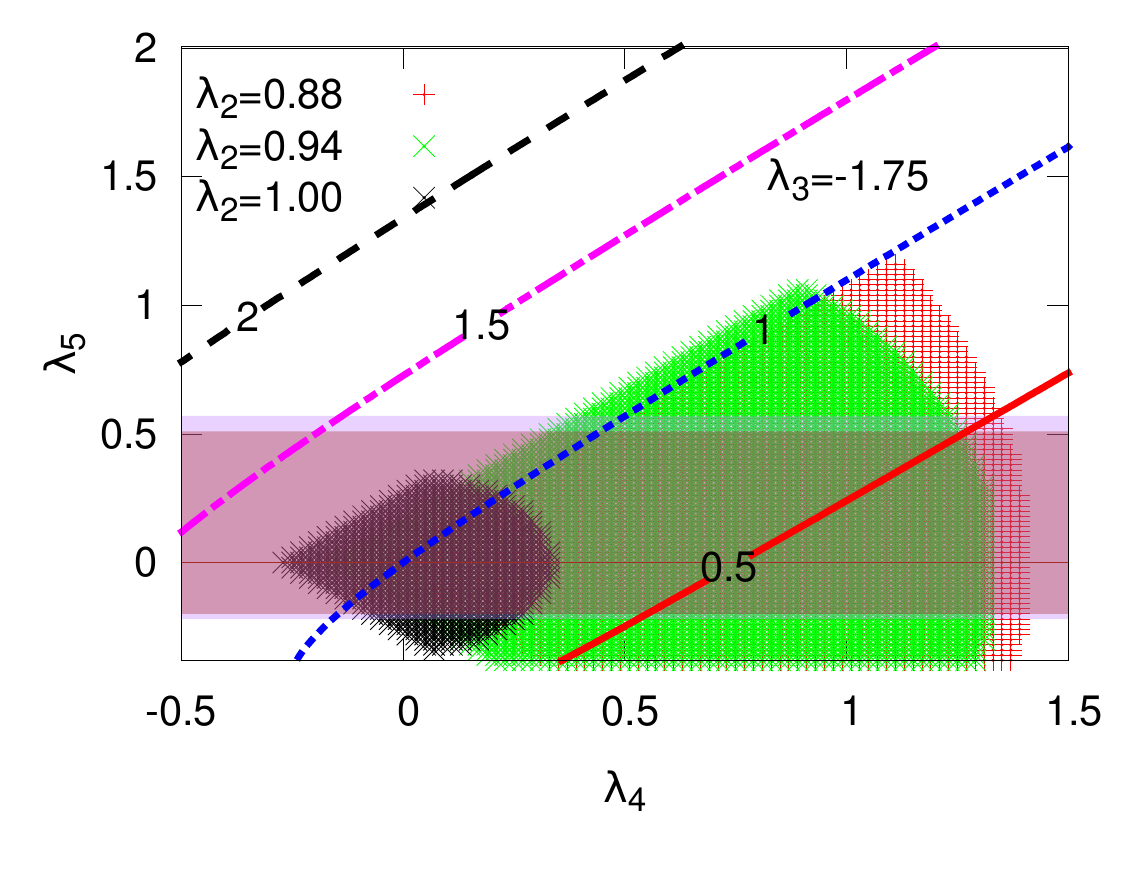}\hspace{-0.2cm}
\includegraphics[scale=0.48]{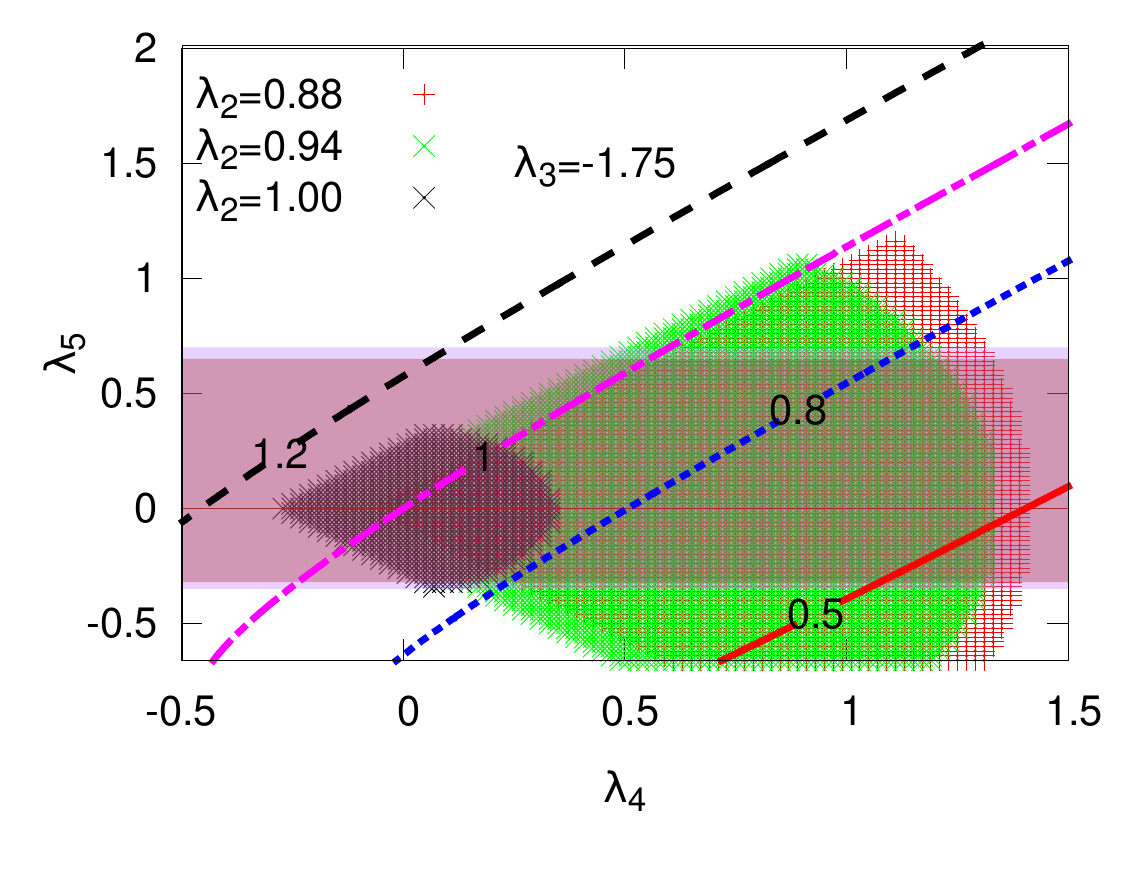}

\hspace{-1.4cm}
\includegraphics[scale=0.48]{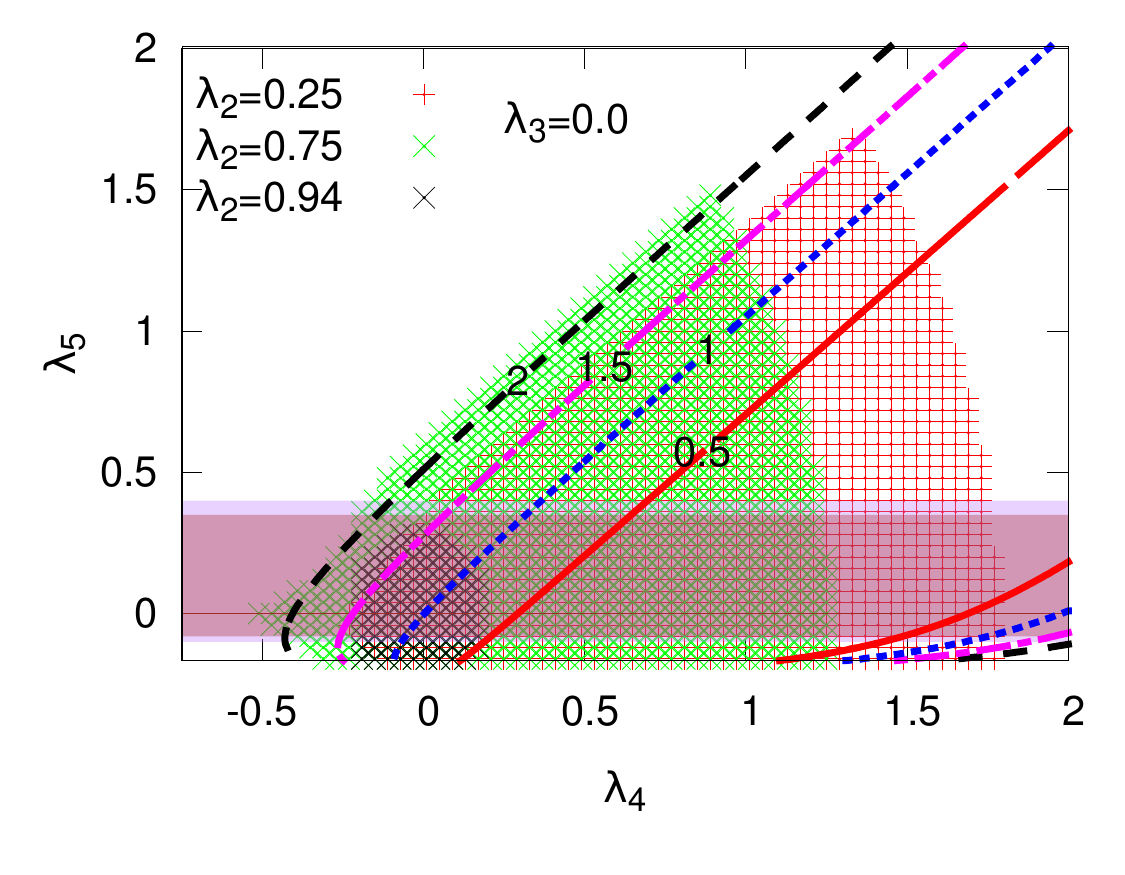}\hspace{-0.2cm}
\includegraphics[scale=0.48]{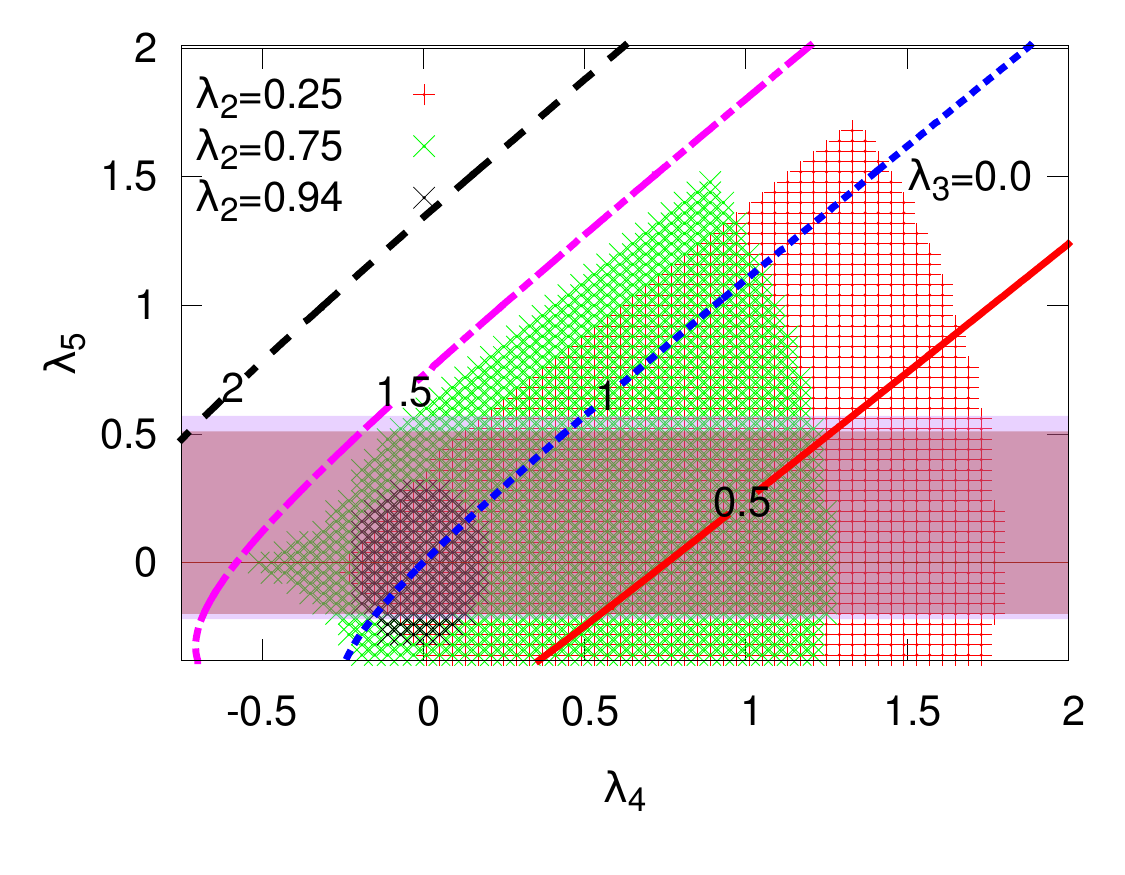}\hspace{-0.2cm}
\includegraphics[scale=0.48]{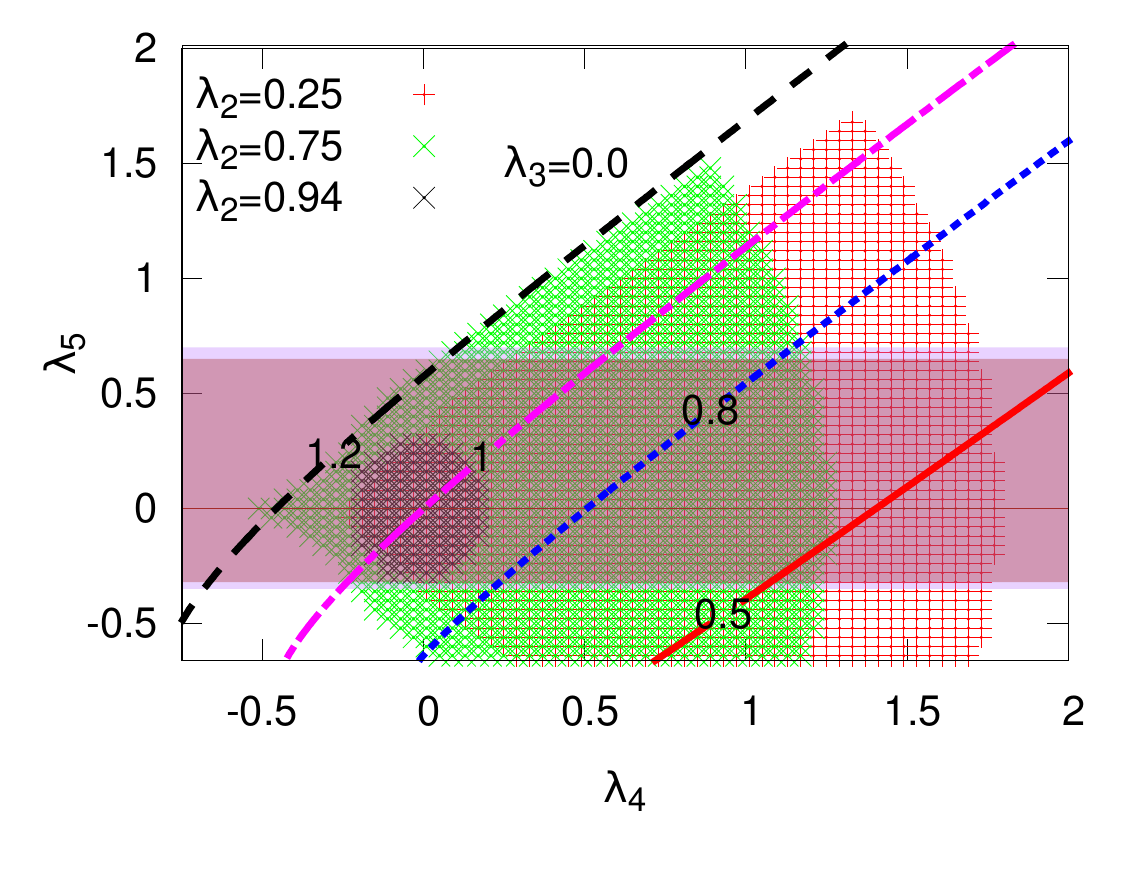}

\hspace{-1.4cm}
\includegraphics[scale=0.48]{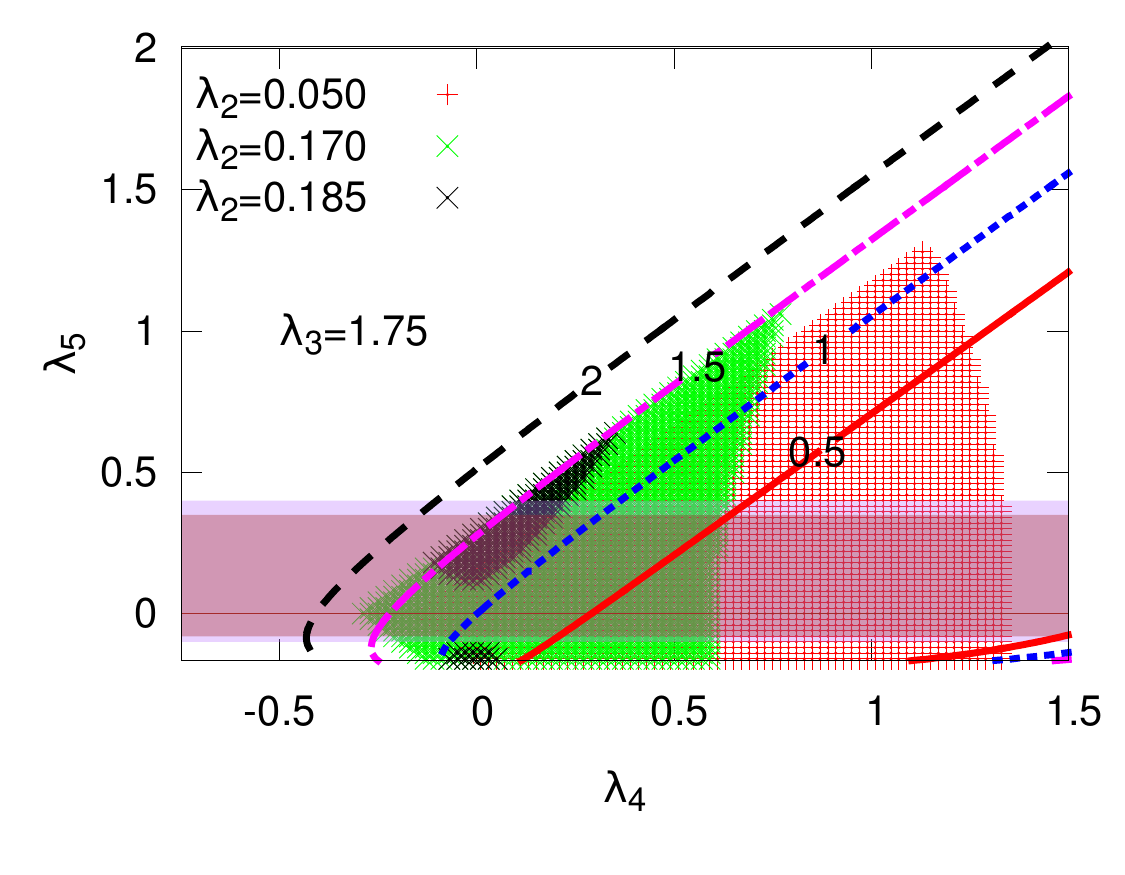}\hspace{-0.2cm}
\includegraphics[scale=0.48]{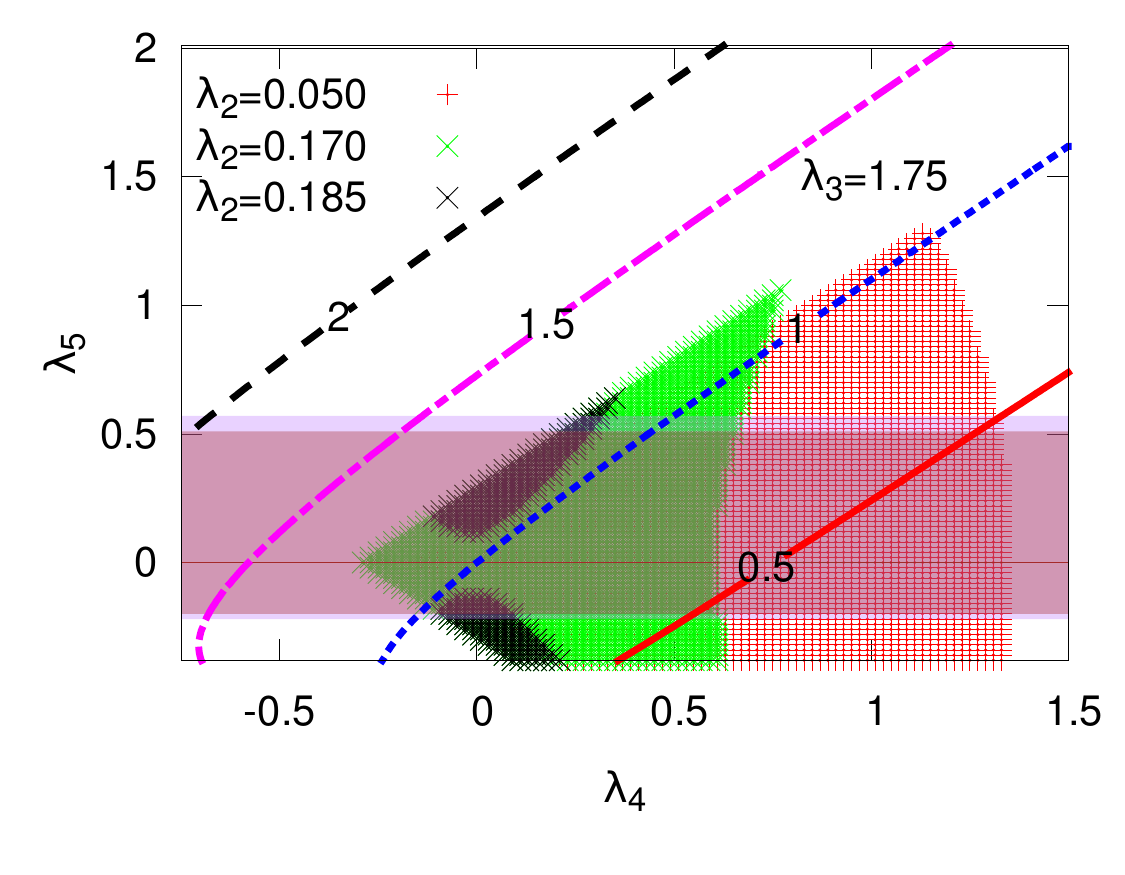}\hspace{-0.2cm}
\includegraphics[scale=0.48]{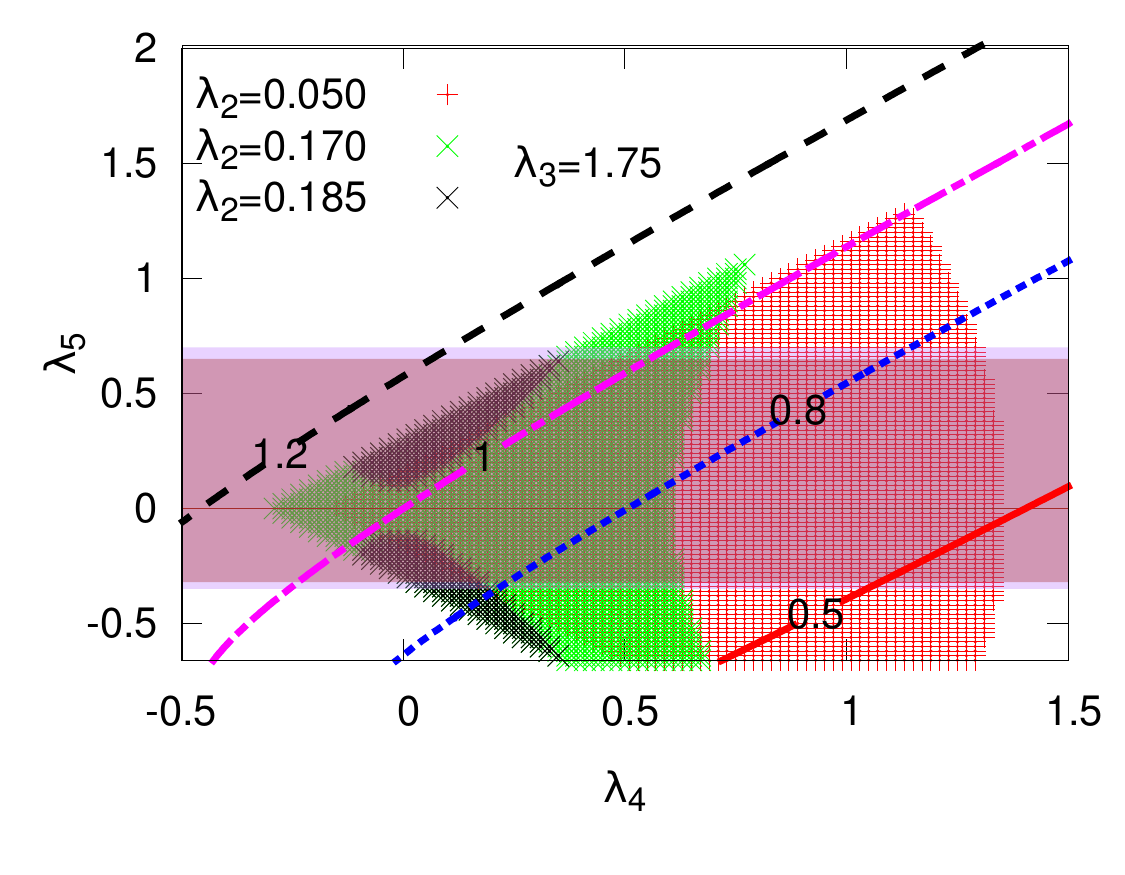}
\end{center}
\caption{Allowed parameter space in the $\lambda_4$--$\lambda_5$ plane
with different values of $\lambda_2$ and $\lambda_3$ for the doubly charged Higgs mass,
$M_{H^{++}}=100$ GeV (left), 150 GeV (middle) and 200 GeV (right). The contours represent
the values of $R_{\gamma\gamma}$. The gray (purple)
bands denote the 99\% (95\% CL) region satisfying the
EWPD constraints.
The cut-off scale is assumed to be $10^{5}$ GeV.} \label{Cut5}
% \label{lam3:-1.75-5} \label{lam3:0.0-5} \label{lam3:1.75-5}
\end{figure}

\begin{figure}[t]
\begin{center}

\hspace{-1.4cm}
\includegraphics[scale=0.48]{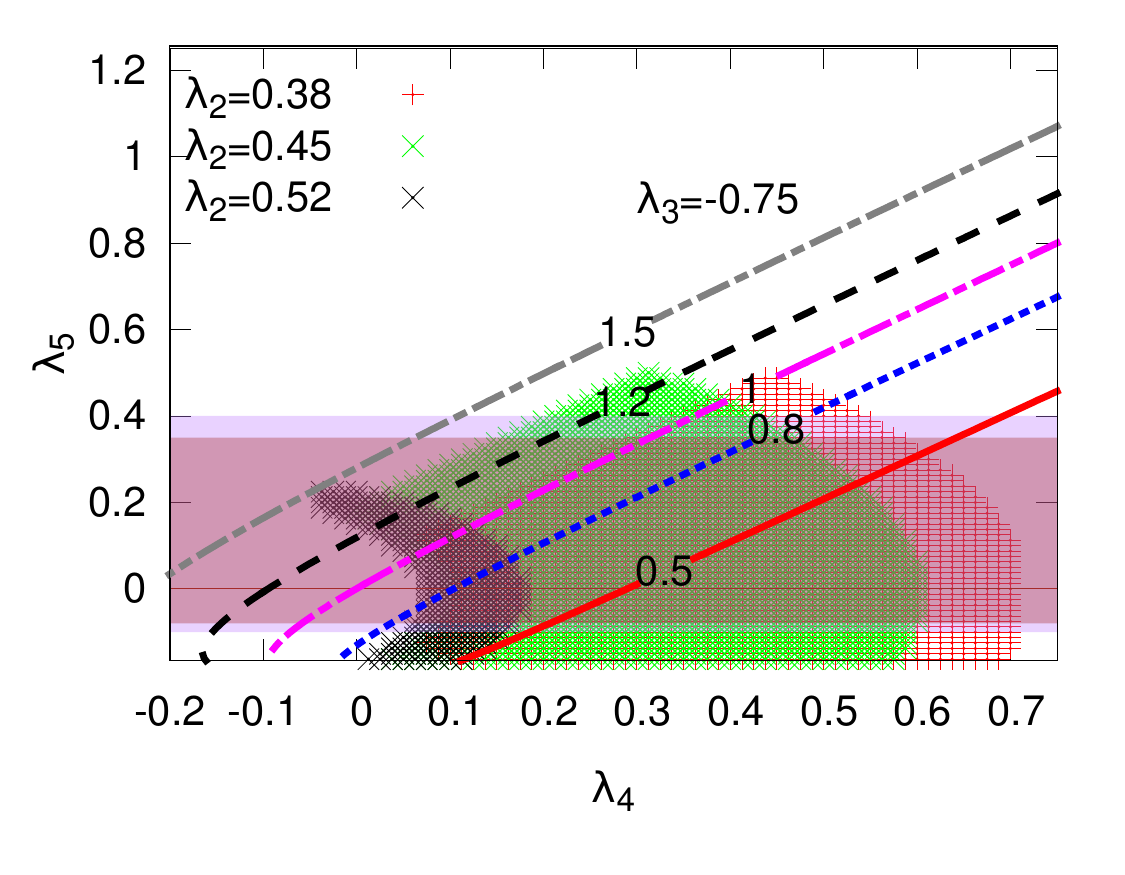}\hspace{-0.2cm}
\includegraphics[scale=0.48]{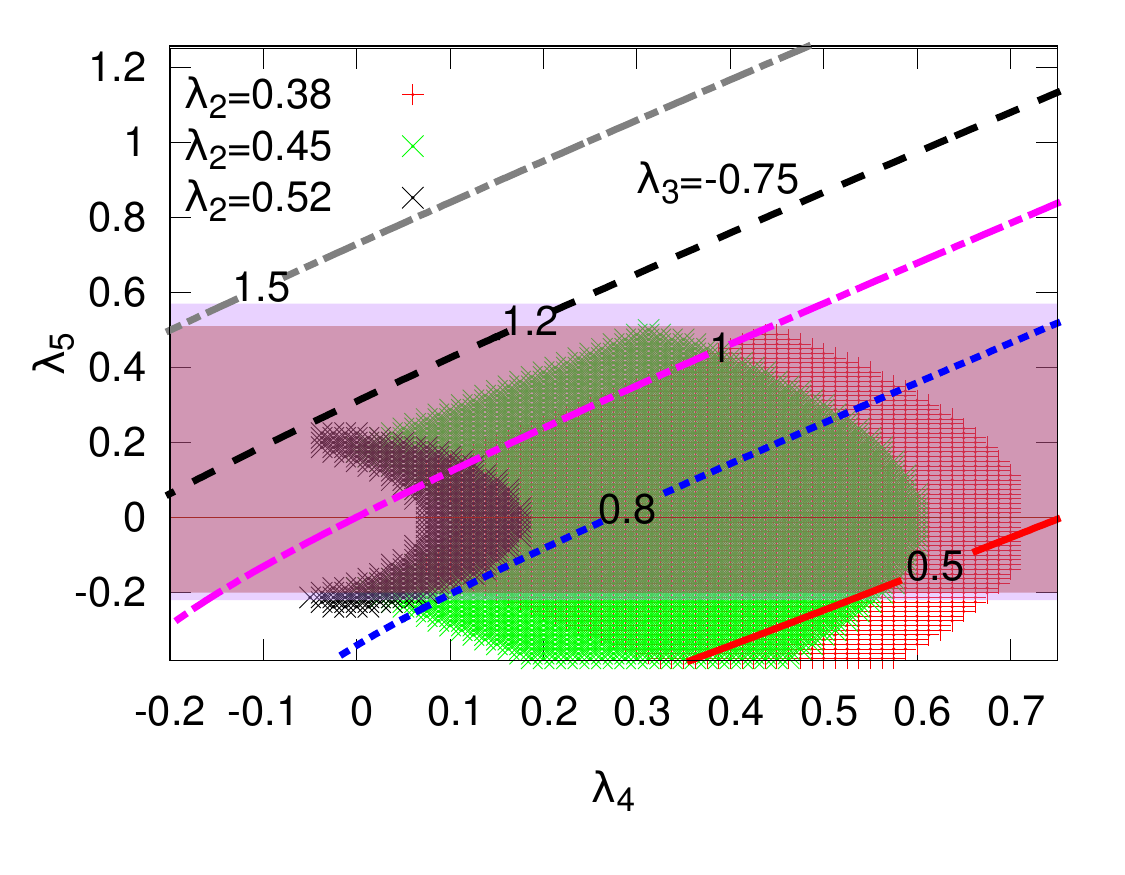}\hspace{-0.2cm}
\includegraphics[scale=0.48]{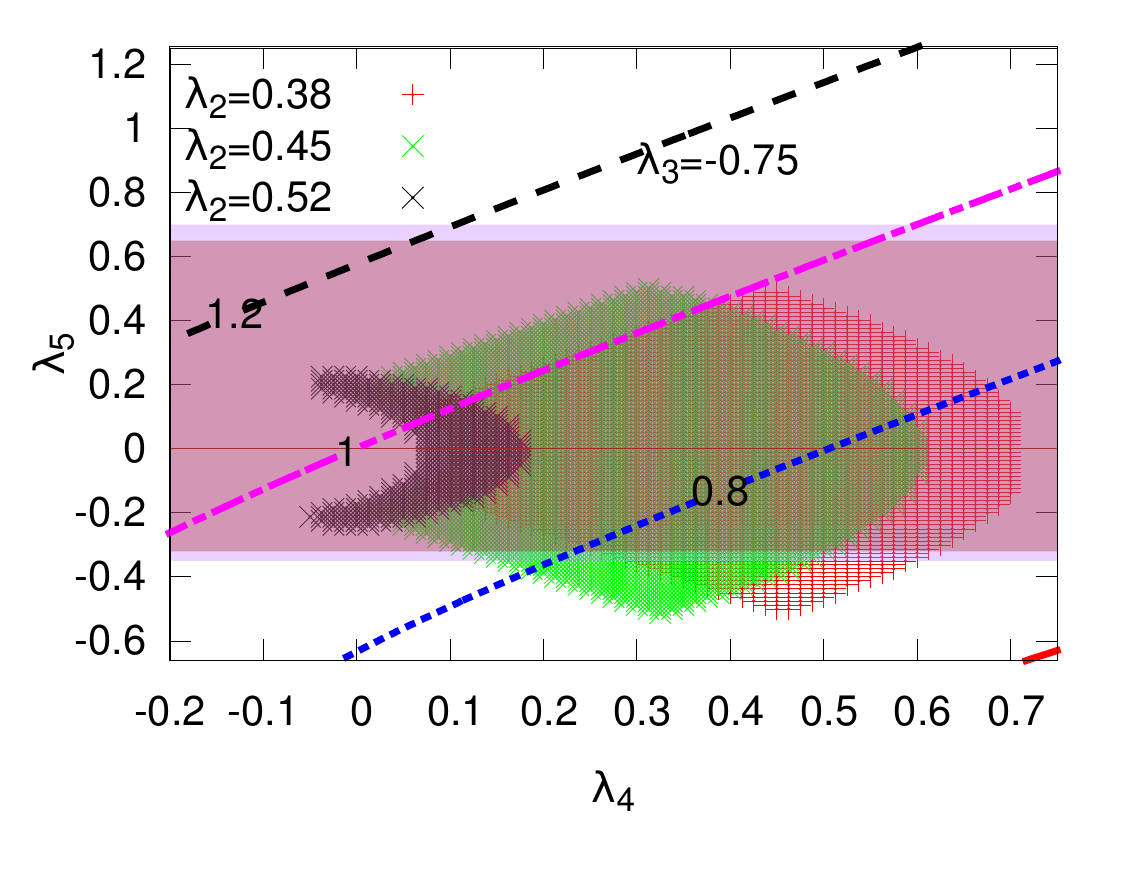}

\hspace{-1.4cm}
\includegraphics[scale=0.48]{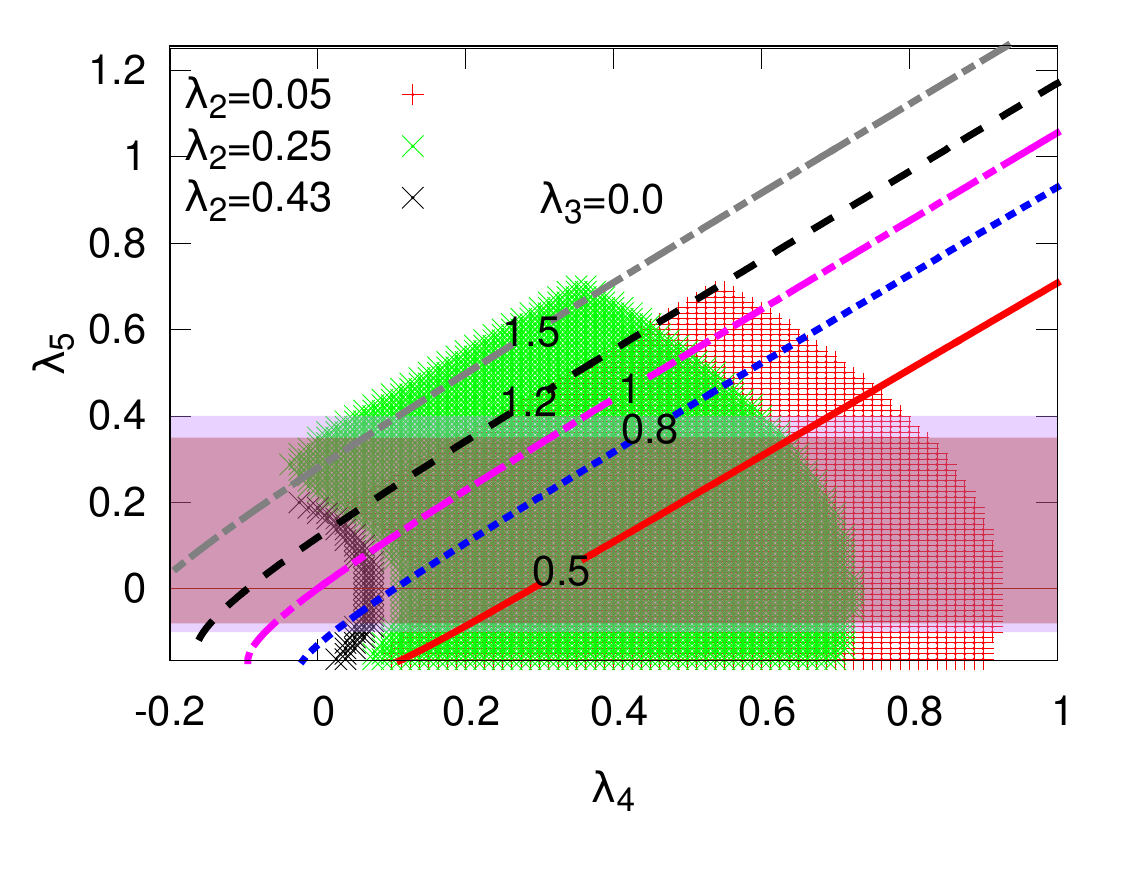}\hspace{-0.2cm}
\includegraphics[scale=0.48]{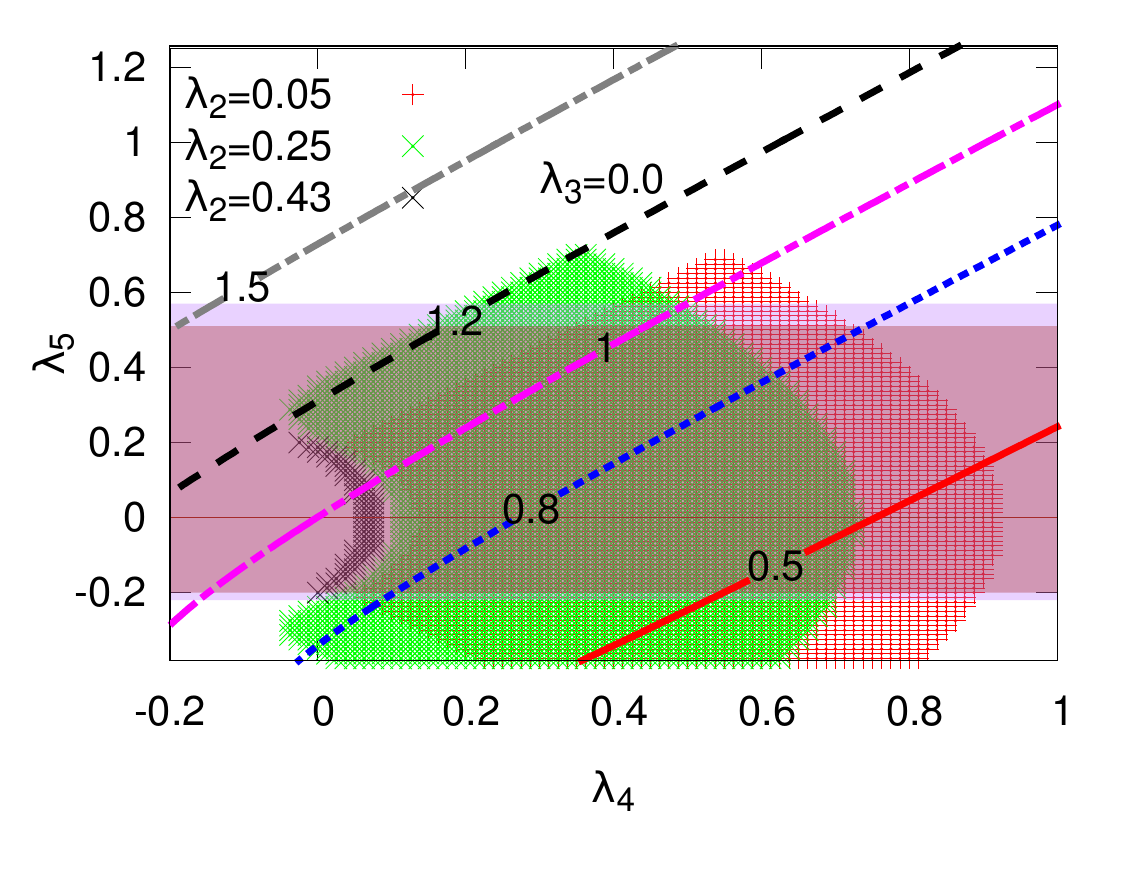}\hspace{-0.2cm}
\includegraphics[scale=0.48]{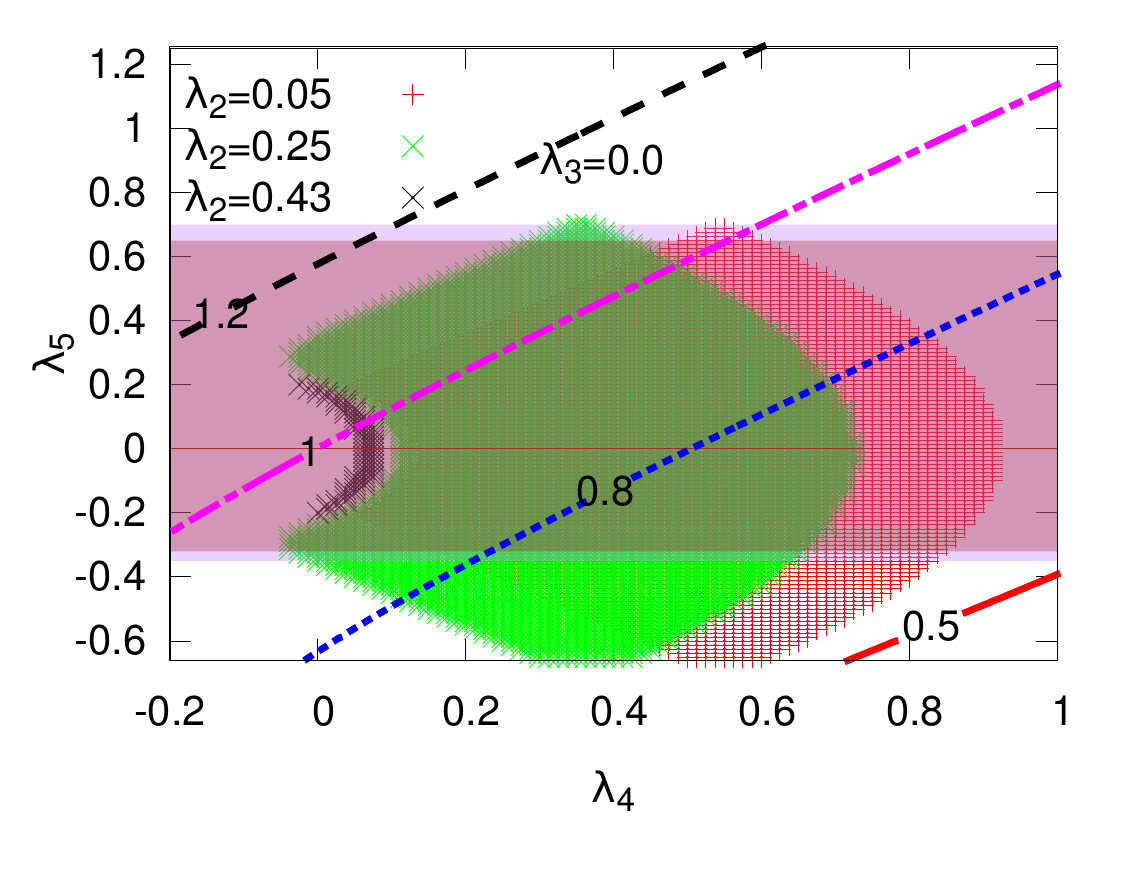}

\hspace{-1.4cm}
\includegraphics[scale=0.48]{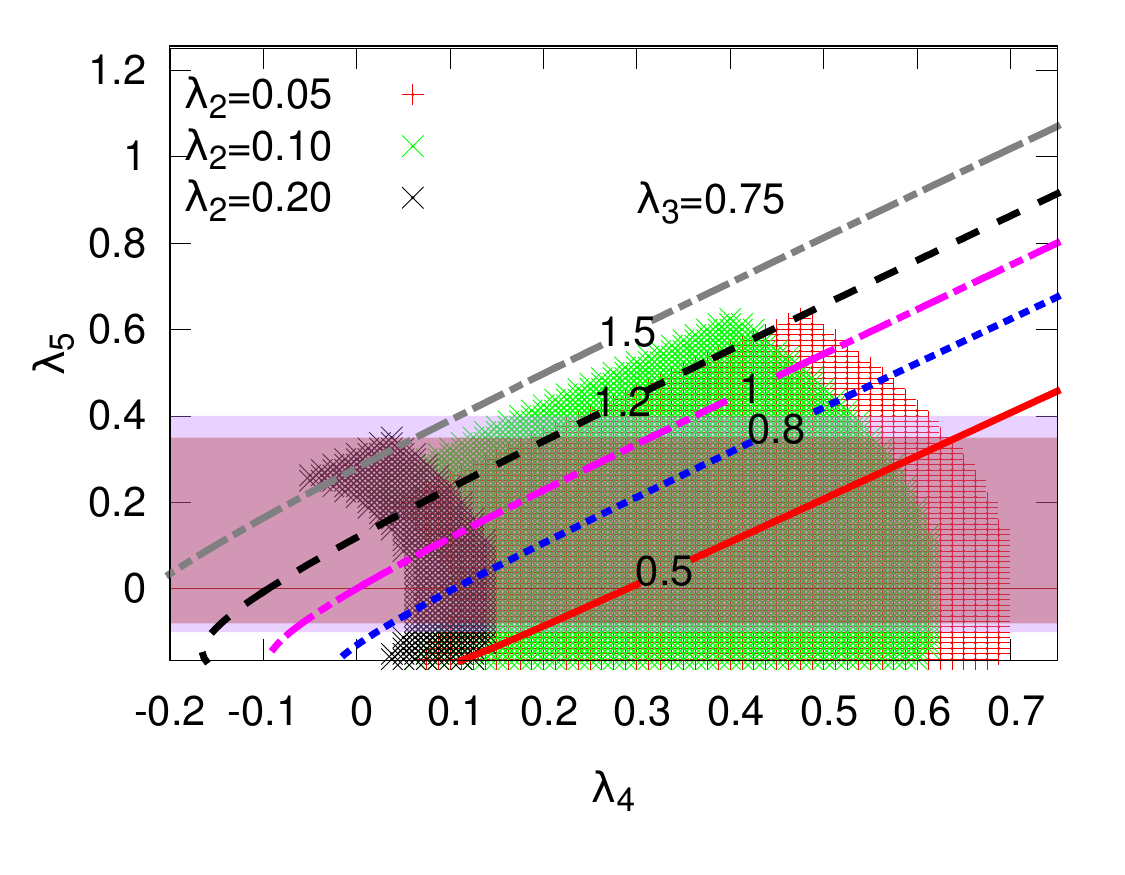}\hspace{-0.2cm}
\includegraphics[scale=0.48]{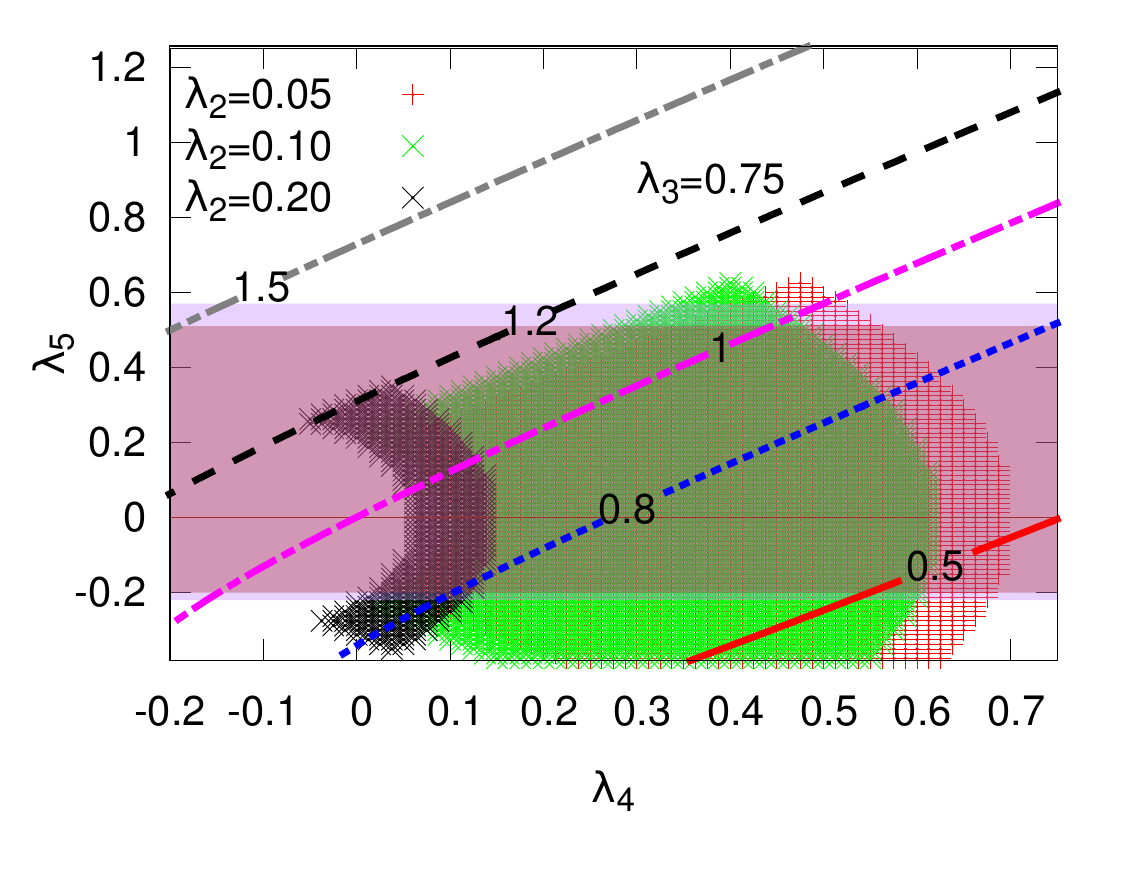}\hspace{-0.2cm}
\includegraphics[scale=0.48]{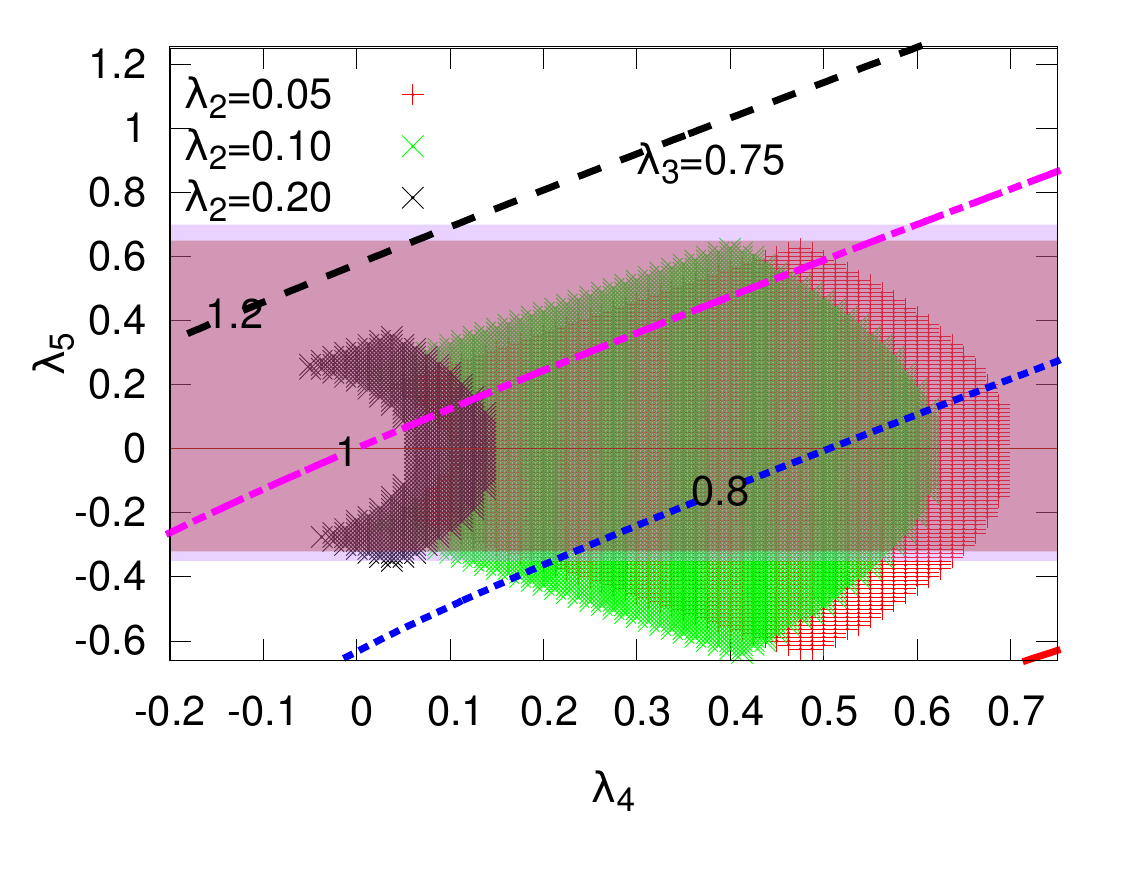}

\end{center}
\caption{The same as Fig.~4, apart from that the cut-off scale is assumed to be $10^{10}$ GeV.} \label{Cut10}
% \label{lam3:-0.75-10} \label{lam3:0.0-10} \label{lam3:0.75-10}
\end{figure}

\begin{figure}[t]
\begin{center}

\hspace{-1.4cm}
\includegraphics[scale=0.48]{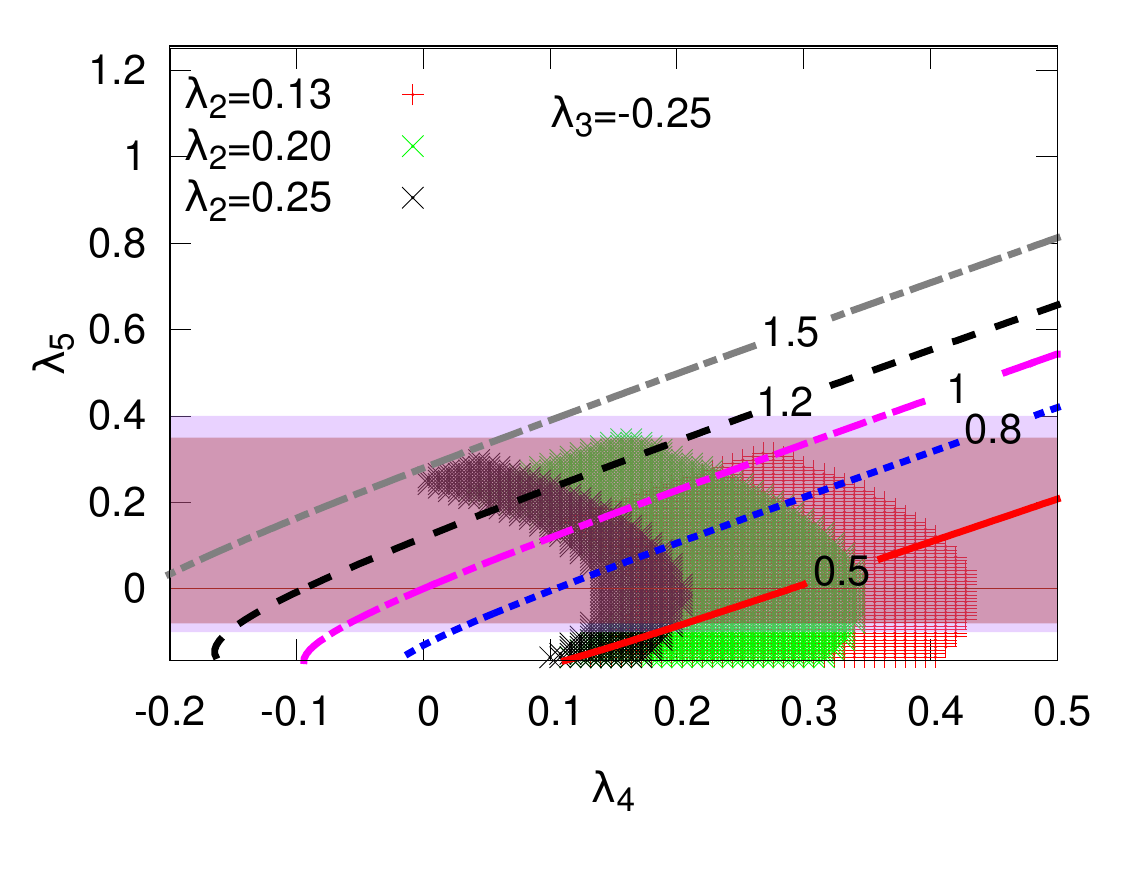}\hspace{-0.2cm}
\includegraphics[scale=0.48]{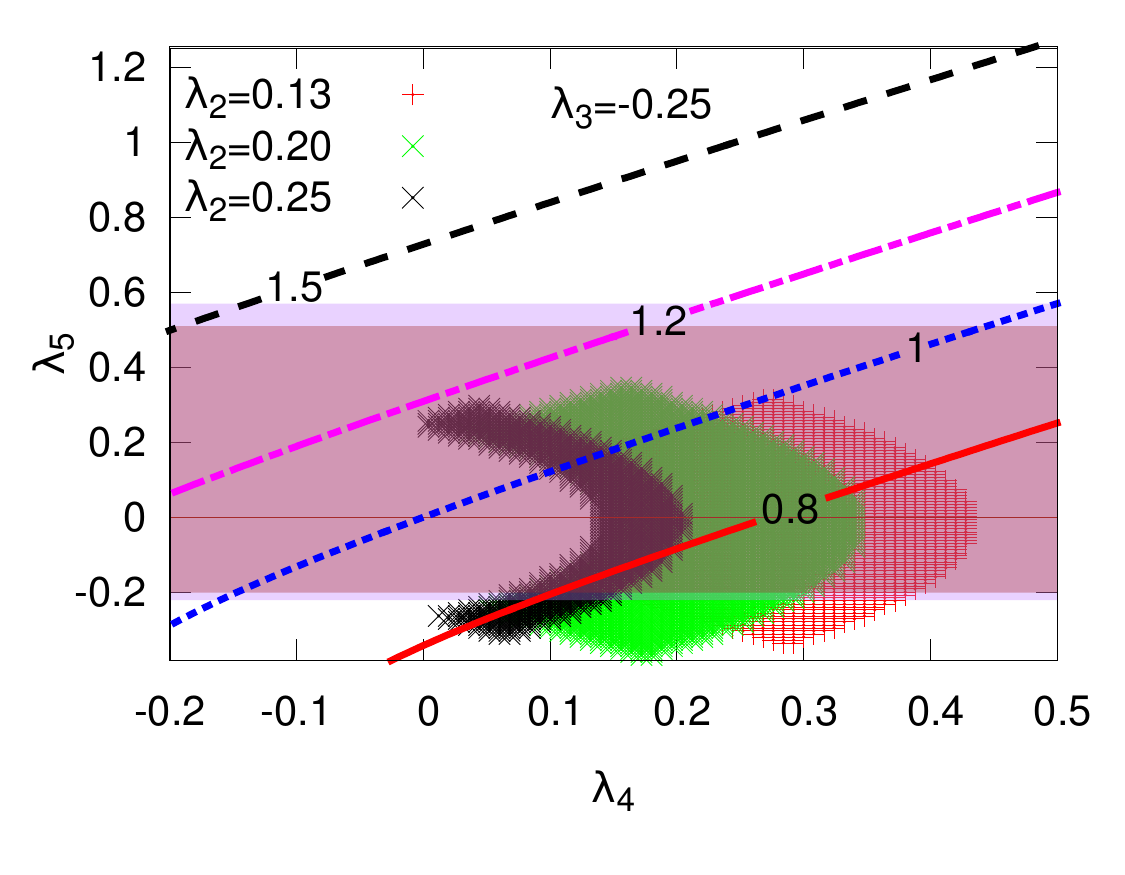}\hspace{-0.2cm}
\includegraphics[scale=0.48]{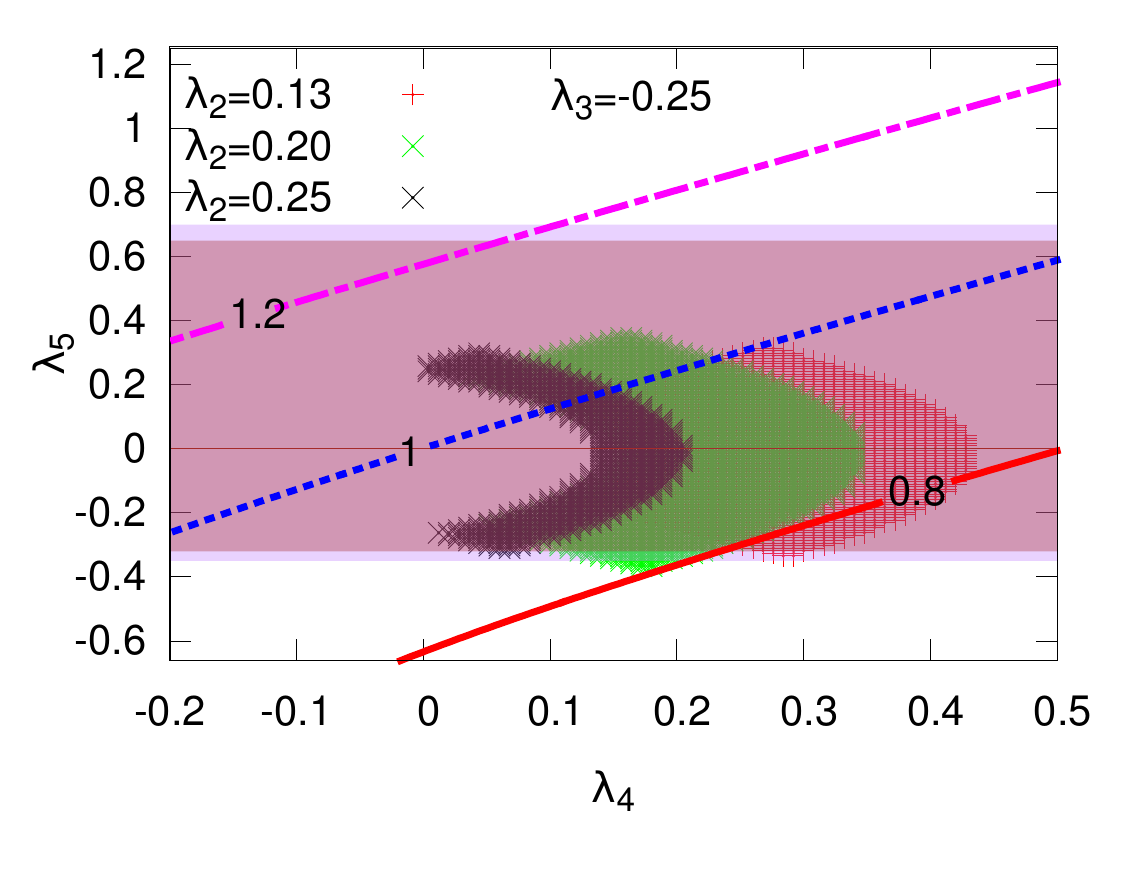}

\hspace{-1.4cm}
\includegraphics[scale=0.48]{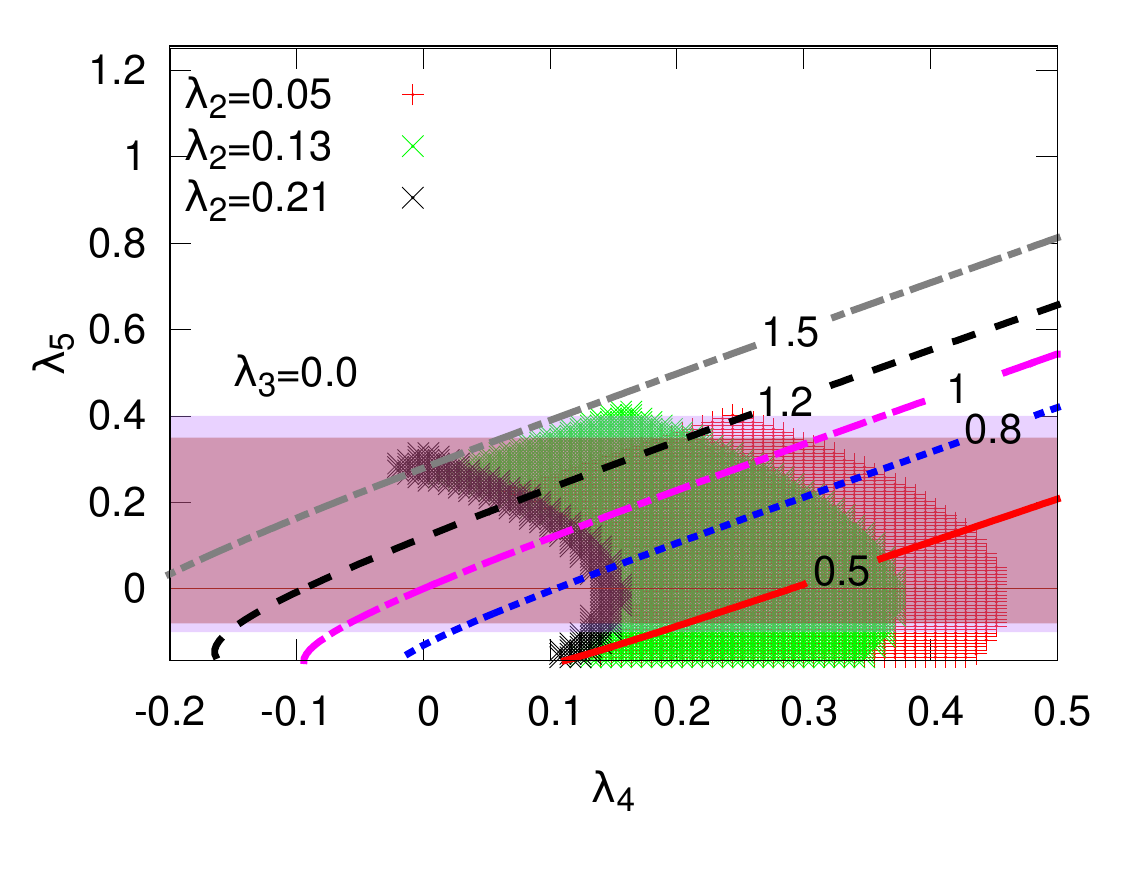}\hspace{-0.2cm}
\includegraphics[scale=0.48]{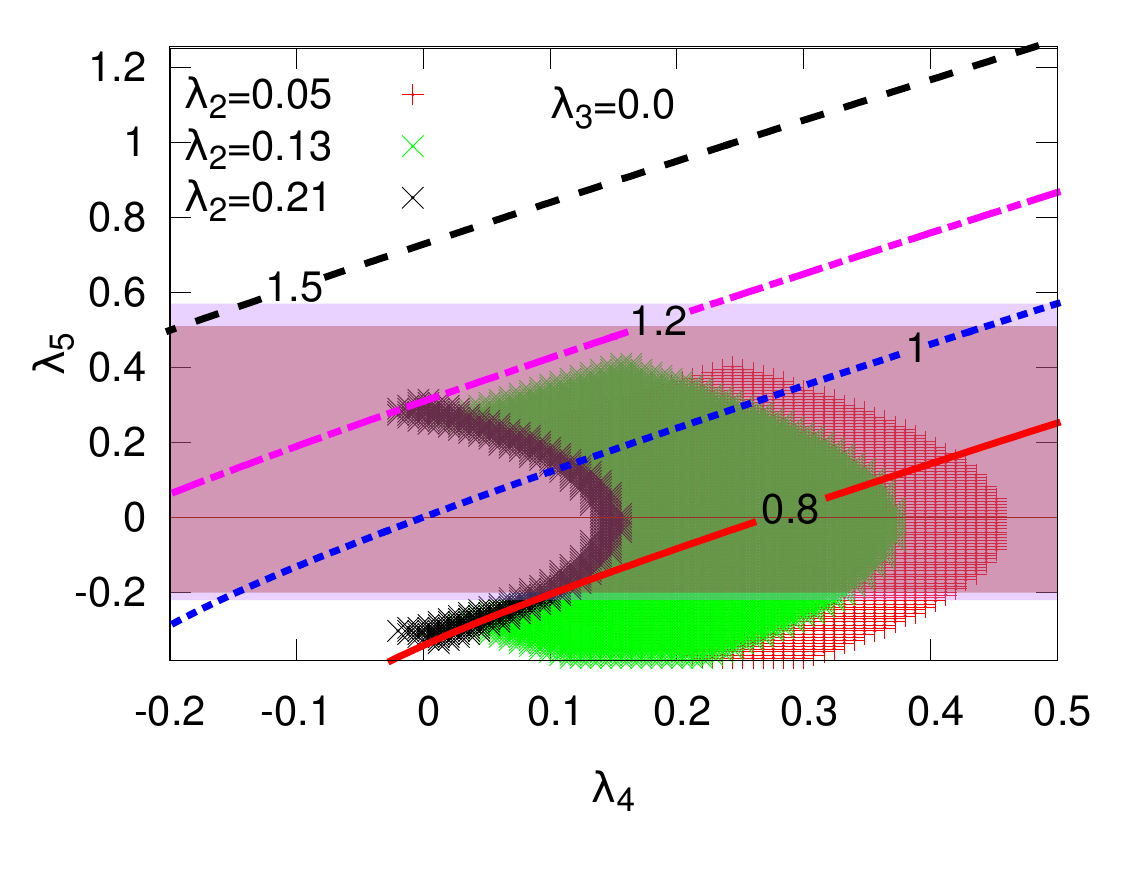}\hspace{-0.2cm}
\includegraphics[scale=0.48]{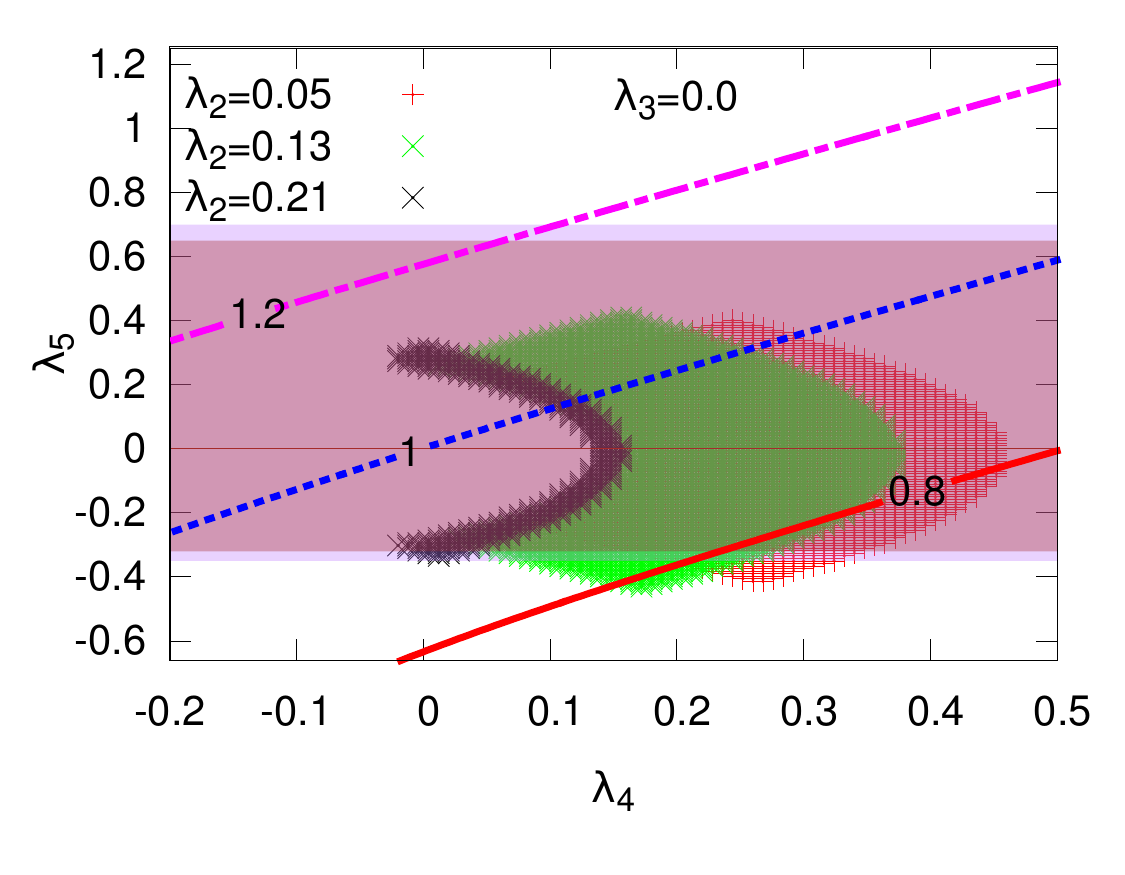}

\hspace{-1.4cm}
\includegraphics[scale=0.48]{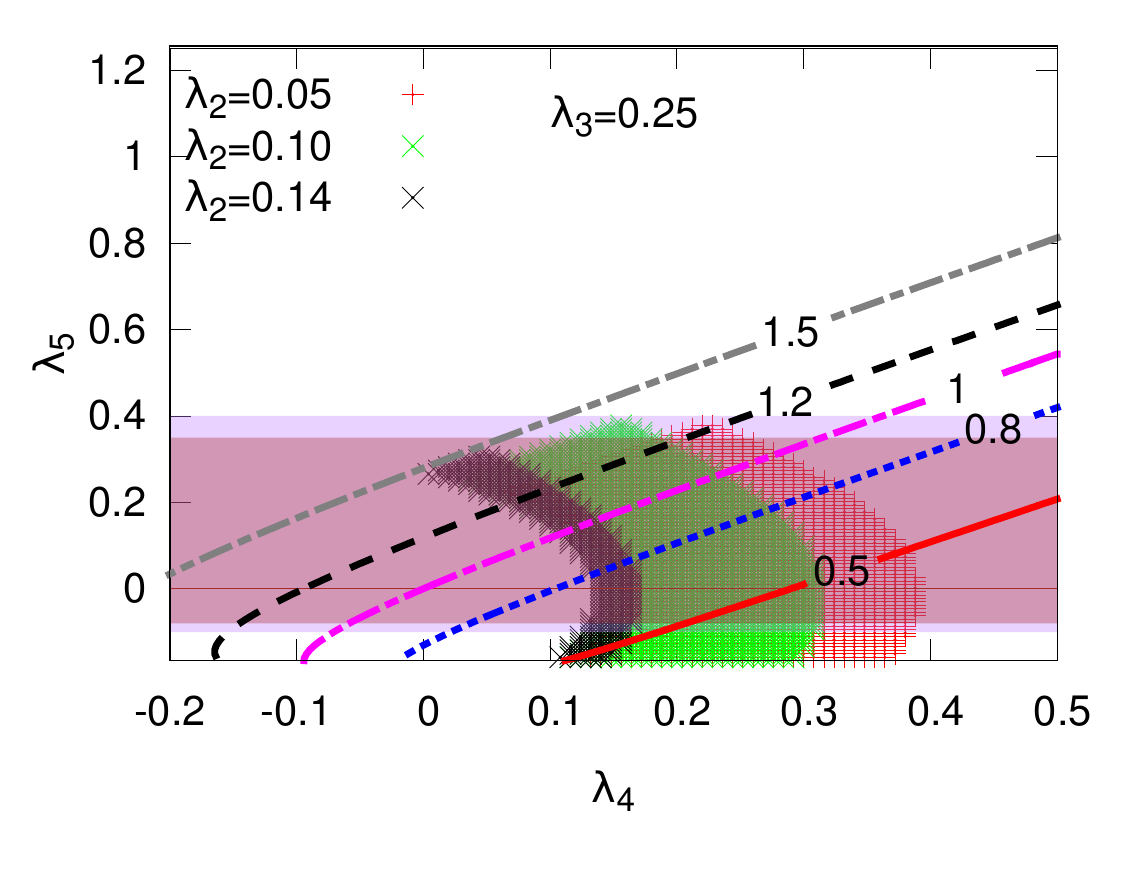}\hspace{-0.2cm}
\includegraphics[scale=0.48]{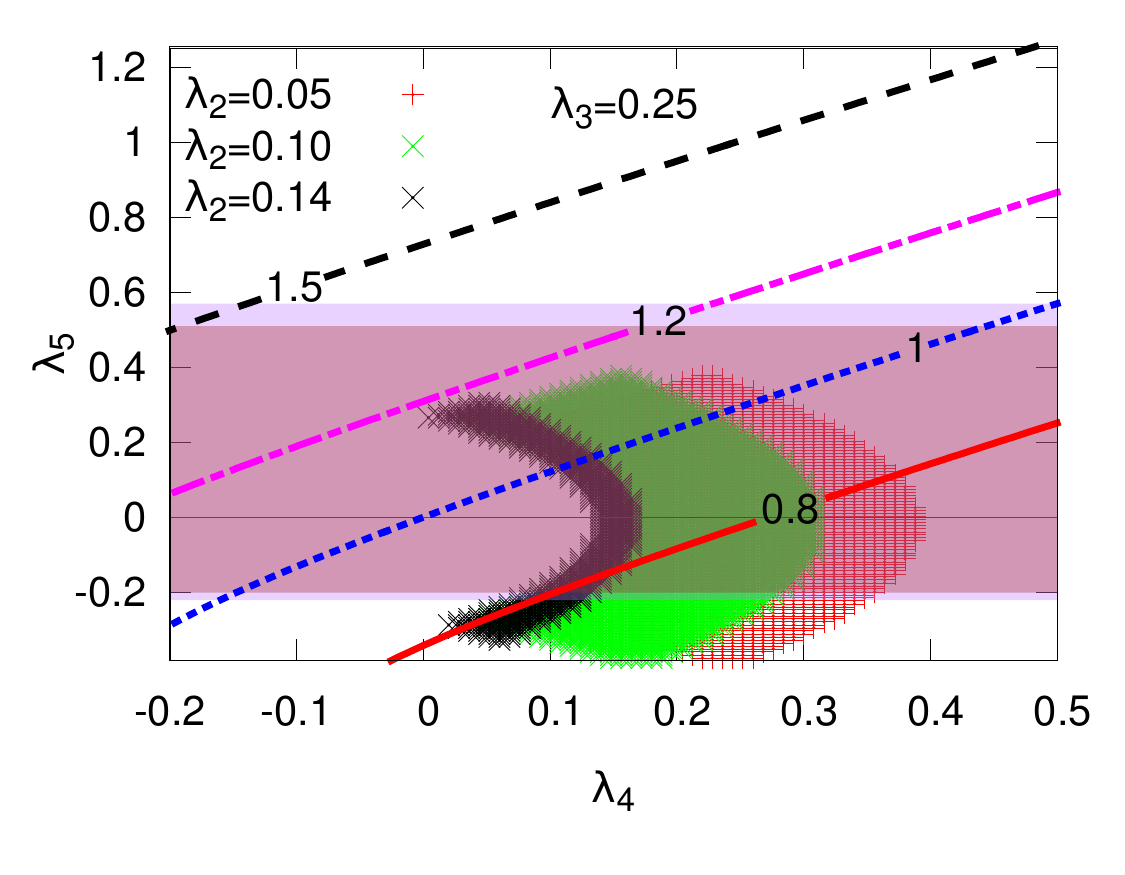}\hspace{-0.2cm}
\includegraphics[scale=0.48]{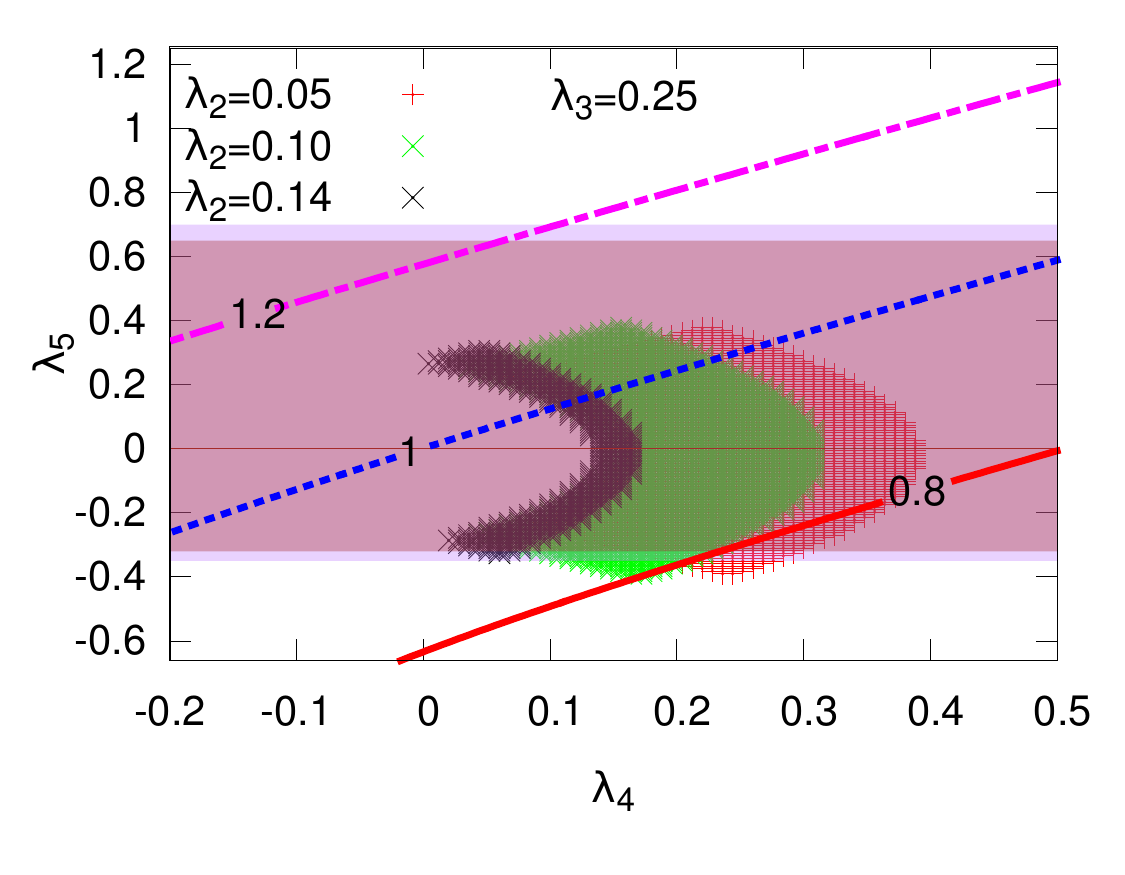}
\end{center}
\caption{The same as Fig.~4, apart from that the cut-off scale is assumed to be $10^{19}$ GeV.} \label{Cut19}
% \label{lam3:-0.25-19}\label{lam3:0.0-19}\label{lam3:0.25-19}
\end{figure}

In this section we perform a numerical analysis to constrain
the parameter space of the scalar couplings by considering
the conditions of vacuum stability and perturbativity
up to the scale where the theory is considered to be valid. We present our
results for three instability scales: 100 TeV, $10^{10}$ GeV and $10^{19}$ GeV in Fig.~\ref{Cut5}, \ref{Cut10}
and \ref{Cut19}, respectively. We further look for
the allowed parameter space combining these with the EWPD and quantify the deviation of the ratio
$R_{\gamma\gamma}$ from the SM value $R_{\gamma\gamma}^{SM}=1$.
%
%In our numerical analysis, we consider We use $M_t = 173$ GeV, $m_t(M_t)=164$ GeV,
%$m_h=125$ GeV, $\lambda_1(M_t)= m_h^2/2 v^2=0.126$ and $y_t(M_t)=\sqrt{2} m_t/v=0.938$.
%
Figs.~\ref{Cut5}--\ref{Cut19} summarize our results in the $\lambda_4$--$\lambda_5$ plane
with different values of $\lambda_2$ and $\lambda_3$ for the doubly charged Higgs mass,
$M_{H^{++}}=100$ GeV (left), 150 GeV (middle) and 200 GeV (right). The contours represent
the values of $R_{\gamma\gamma}$. The gray (purple)
bands denote the 99\% (95\% CL) region satisfying the EWPD constraints.

It is obvious from the Figs.~\ref{Cut5}, \ref{Cut10} and \ref{Cut19} that the small cut-off scale
allows a large parameter space while the larger cut-off scale constrains it.
The allowed range of $\lambda_2$ depends upon the range of $\lambda_3$. Disregarding such a correlation,
the maximally allowed ranges of the couplings depending on the cut-off scale are found to be as follows:
\begin{equation}
\begin{array}{c|c|c|c}
           & 10^5 \mbox{ GeV} & 10^{10} \mbox{ GeV} & 10^{19} \mbox{ GeV} \cr
           \hline
\lambda_2 \; & (0, 1) & (0, 0.5) & (0, 0.25)  \cr
\lambda_3\; & ~(-2.0, 2.4)~ & ~(-1.0, 1.25)~ & ~(-0.55,0.62)~ \cr
\lambda_4\; & (-0.5, 1.7) & (-0.1, 0.9) & (0, 0.5) \cr
\lambda_5\; & (-1.5, 1.5) & (-0.7, 0.7) & (-0.4, 0.4) \cr
\end{array}
\end{equation}
The EWPD requiring the triplet mass splitting $|\Delta M| \lesssim 40 $ GeV, allow the following ranges of $\lambda_5$
\begin{equation}
\lambda_5 = (-0.1,0.4),\quad (-0.2,0.6), \quad (-0.35,0.7)
\end{equation}
for $M_{H^{++}}=$ 100, 150, and 200 GeV, respectively.

We infer from these figures that for negative $\lambda_3$, larger values of $\lambda_2$
are allowed while for positive $\lambda_3$, smaller values of $\lambda_2$ are preferred to satisfy vacuum stability conditions.
We observe that a large $\lambda_2$ tends to
squeeze the allowed parameter space in the $\lambda_4$--$\lambda_5$ plane. This is due to the
fact that a large $\lambda_2$ violates perturbativity very quickly
when we evolve the coupling with RG equations.
We find that $\lambda_3=0$ allows for a larger parameter space
compared to two extremal values of $\lambda_3$. As a result, the enhancement of
$R_{\gamma\gamma}$ is feasible for relatively larger allowed parameter space.
The shaded bands in figures denote
the allowed region by the EWPD depending on the doubly charged Higgs boson mass.
As is obvious, smaller and more positive ranges of $\lambda_5$ are
allowed for smaller values of $M_{H^{++}}$.
 Although the allowed bands of $\lambda_5$ get smaller for smaller $M_{H^{++}}$,
 $R_{\gamma\gamma}$ can be more enhanced in these regions due to the sizable contribution from light
charged Higgs bosons, in particular, near $\lambda_4=0$ favored by vacuum stability conditions.
 In the case of $M_{H^{++}}=100 \,(200)$ GeV,
 one can get  $R_{\gamma\gamma}$ as large as 2 (1.2), or 1.5 (1.1)
 for $\Lambda=10^5$, or  $10^{10}$ and $10^{19}$ GeV.
 Of course, a larger parameter space opens up for a larger positive $\lambda_4$
for which a destructive interference occurs and thus $R_{\gamma\gamma}$ can be much smaller than 1.
Thus, broad ranges with positive $\lambda_4$ are strongly disfavored by the current LHC data.

To summarize, we studied the parameter space of the Higgs scalar potential of the type II
seesaw model in the light of vacuum stability, perturbativity and EWPD constraints.
Then we looked at the possible deviation in the Higgs-to-diphoton rate
in the allowed parameter space.
The allowed parameter space is found to be very restrictive and
strongly depend on the choice of the instability scale.
Regardless of any choice of instability scale, $R_{\gamma\gamma}$ becomes smaller than 1 in a larger
parameter space,  but it can be enhanced by 50\%-100\% in some limited parameter region.
 If the deviation of the Higgs-to-diphoton rate turns out to be small with more data
at the LHC, only a narrow band around $\lambda_4 \approx \lambda_5$ will survive
for low Higgs triplet mass.

\acknowledgments
EJC was supported by the National Research Foundation of Korea (NRF) grant funded by the Korea government (MEST) (No.~20120001177).

\end{document}